\documentclass[iop,numberedappendix,twocolappendix]{emulateapj}
\usepackage[breaklinks,colorlinks,citecolor=blue]{hyperref}
\usepackage{amsmath}
\usepackage{multirow}
\usepackage{color}

\newcommand{\ha}{H$\alpha$}

\shorttitle{UVUDF UV LFs}
\shortauthors{Mehta et al.}
\slugcomment{{\sc Accepted to \apj:} February 20, 2017} 

\begin{document}

\title{UVUDF: UV Luminosity Functions at the cosmic high-noon}
\author{Vihang Mehta\altaffilmark{1},
Claudia Scarlata\altaffilmark{1}, 
Marc Rafelski\altaffilmark{2,3,11},
Timothy Gburek\altaffilmark{4},
Harry I. Teplitz\altaffilmark{5},
Anahita Alavi\altaffilmark{4},
Michael Boylan-Kolchin\altaffilmark{6},
Steven Finkelstein\altaffilmark{6},
Jonathan P. Gardner\altaffilmark{3},
Norman Grogin\altaffilmark{2},
Anton Koekemoer\altaffilmark{2},
Peter Kurczynski\altaffilmark{7},
Brian Siana\altaffilmark{4},
Alex Codoreanu\altaffilmark{8},
Duilia F. de Mello\altaffilmark{3,9},
Kyoung-Soo Lee\altaffilmark{10},
Emmaris Soto\altaffilmark{9}}

\altaffiltext{1}{Minnesota Institute for Astrophysics, University of Minnesota, Minneapolis, MN 55455, USA}
\altaffiltext{2}{Space Telescope Science Institute, Baltimore, MD 21218, USA}
\altaffiltext{3}{Goddard Space Flight Center, Code 665, Greenbelt, MD, 20771, USA}
\altaffiltext{4}{Department of Physics and Astronomy, University of California Riverside, Riverside, CA 92521, USA}
\altaffiltext{5}{Infrared Processing and Analysis Center, Caltech, Pasadena, CA 91125, USA}
\altaffiltext{6}{Department of Astronomy, The University of Texas at Austin, Austin, TX 78712, USA}
\altaffiltext{7}{Department of Physics and Astronomy, Rutgers, The State University of New Jersey, Piscataway, NJ 08854, USA}
\altaffiltext{8}{Centre for Astrophysics and Supercomputing, Swinburne University of Technology, Melbourne, Australia}
\altaffiltext{9}{The Catholic University of America, Physics Department, Washington DC, 20064}
\altaffiltext{10}{Department of Physics, Purdue University, West Lafayette, IN 47906}
\altaffiltext{11}{NASA Postdoctoral Fellow}

\begin{abstract}
We present the rest-1500\AA\ UV luminosity functions (LF) for star-forming galaxies during the cosmic \textit{high noon} -- the peak of cosmic star formation rate at $1.5<z<3$. We use deep NUV imaging data obtained as part of the \textit{Hubble} Ultra-Violet Ultra Deep Field (UVUDF) program, along with existing deep optical and NIR coverage on the HUDF. We select F225W, F275W and F336W dropout samples using the Lyman break technique, along with samples in the corresponding redshift ranges selected using photometric redshifts and measure the rest-frame UV LF at $z\sim1.7,2.2,3.0$ respectively, using the modified maximum likelihood estimator. We perform simulations to quantify the survey and sample incompleteness for the UVUDF samples to correct the effective volume calculations for the LF. We select galaxies down to $M_{UV}=-15.9,-16.3,-16.8$ and fit a faint-end slope of $\alpha=-1.20^{+0.10}_{-0.13}, -1.32^{+0.10}_{-0.14}, -1.39^{+0.08}_{-0.12}$ at $1.4<z<1.9$, $1.8<z<2.6$, and $2.4<z<3.6$, respectively. We compare the star formation properties of $z\sim2$ galaxies from these UV observations with results from \ha\ and UV$+$IR observations. We find a lack of high SFR sources in the UV LF compared to the \ha\ and UV$+$IR, likely due to dusty SFGs not being properly accounted for by the generic $IRX-\beta$ relation used to correct for dust. We compute a volume-averaged UV-to-\ha\ ratio by \textit{abundance matching} the rest-frame UV LF and \ha\ LF. We find an increasing UV-to-\ha\ ratio towards low mass galaxies ($M_\star \lesssim 5\times10^9$ M$_\odot$). We conclude that this could be due to a larger contribution from starbursting galaxies compared to the high-mass end.
\end{abstract}

\section{Introduction}
\label{sec:intro}

The galaxy luminosity function is one of the key observables in astronomy, providing the number density of galaxies at a given luminosity and time. The luminosity function is instrumental in establishing the connection between the observable light and the underlying distribution of dark matter halos. The link between these two depends on the baryonic physics which ultimately regulate the conversion of gas into stars and the luminosity output at any given wavelength.

In the rest--frame ultra violet (UV), the galaxy continuum is  dominated by light coming from young stars, and is therefore a direct tracer of recent star formation activity. Consequently, the UV luminosity function can be used to describe the volume averaged star formation rate in the Universe and to study the in-situ build up of stellar mass in galaxies. Moreover, unlike other star formation indicators, the rest-frame UV is continuously accessible to very high redshifts and hence is an invaluable diagnostic for mapping star formation out to very early times.

A large amount of effort has been devoted into obtaining accurate measurements of the rest-frame UV luminosity function at all redshifts $z \lesssim 10$ \citep[e.g.,][]{arnouts05, sawicki06, yoshida06, bouwens07, dahlen07, reddy09, hathi10, oesch10, vanderburg10, sawicki12, alavi14, alavi16, bouwens14a, bouwens14b, bouwens15, finkelstein15, bernard16, parsa16}. These observations show that the UV luminosity density increases steadily up to $z\sim 2-3$, followed by  a slight decline out to the highest redshifts probed so far \citep[e.g., see][]{alavi16}. 

Recently, very faint galaxies have attracted significant attention for a variety of reasons. At $z\sim6-10$, they are expected to be essential for reionization of the Universe \citep{bouwens12,finkelstein12,jaacks12,robertson13,robertson15}, and they include likely progenitors of L$^\star$ galaxies in the local Universe \citep[e.g.,][]{boylan15}. At intermediate redshifts, faint galaxies provide excellent tests of feedback due to star formation and reionization \citep[e.g.,][]{benson03,lofaro09,weinmann12}. In the nearby Universe, these systems probe galaxy formation on the finest scales and may contain clues to the nature of dark matter \citep[e.g.,][]{menci12,menci16,nierenberg13,kennedy14}. The evolution of the faint-end slope ($\alpha$) of the UV luminosity function can therefore inform us on many crucial aspects of galaxy formation and evolution.

It is not surprising then, that the value of $\alpha$ and its time evolution has been the subject of much research, and it is highly debated in the current literature. From published results, $\alpha$ appears to evolve dramatically, going from $\alpha\sim-1.2$ at $z\sim0$ to $\alpha\sim-2$ by $z\sim8$, albeit with a rather large scatter. At $z\sim2$, faint-end slope estimates vary from considerably shallow values of $\alpha\sim-1.3$ \citep{hathi10,parsa16} to very steep values of $\alpha=-1.72$ \citep{alavi14,alavi16}.  The survey limits are the main challenge in accessing the faint galaxies needed to significantly constrain the value of $\alpha$ \citep[e.g.,][]{oesch10}. Strong gravitational lensing enables one to circumvent this limitation, although, it introduces additional systematics and complications, such as a non-trivial effective survey volume calculation \citep[e.g.,][]{alavi14,alavi16}. Deep, direct imaging still provides the most robust estimate for $\alpha$. 

Complementary to the UV, the \ha\ recombination line is a gold--standard indicator for ongoing star formation. These two tracers, however, are sensitive to star formation occurring over different timescales\footnote{The \ha\ emission traces star formation over short time scales ($\sim$ 10s of Myrs, typical of the hot, O- and B-type stars that power HII regions).  On the other hand, the contribution to the rest-frame UV continuum comes from the longer lived B-A stars ($\sim$100 Myrs).} \citep{kennicutt12}, and are affected differently by interstellar dust attenuation. In the local Universe, the two indicators are found to agree with each other, under the assumption that the star-formation has been constant over a long enough time to allow equilibrium  \citep[$>100$Myr; e.g.,][]{buat87, buat92, bell01, iglesias04, salim07, lee09, barnes11, hermanowicz13}. The effect of dust attenuation in the rest-frame UV is usually corrected using locally-calibrated empirical relations between the slope of the UV continuum and the IR excess \citep[$IRX-\beta$ relation, ][]{meurer99}. The \citet{meurer99} relation was calibrated for central starbursts in the nearby Universe. As a whole, star forming galaxies in the nearby Universe lie below this relation, as found by many studies \citep[e.g.,][]{munoz09, boquien12, grasha13, ye16}.

At high redshifts, it has been suggested that star formation is dominated by more stochastic, intense bursts which may also be more important in low(er)--mass galaxies \citep[e.g.,][]{shen13,hopkins14,dominguez15}. If this is true, the constant star formation history assumption implicit in all luminosity-to-star formation rate conversions breaks down. There are also indications that the \citet{meurer99} correction for dust may not be adequate at $z\gtrsim 1$ \citep[e.g.][]{buat12, dayal12, wilkins12, debarros14, castellano14, smit15, reddy15, talia15, alvarez16}.  Until \textit{JWST} comes online, the highest possible redshift where a direct comparison of the two SFR indicators can be performed is  $z=2.5$. 

In this paper, we use the UVUDF \citep{teplitz13}, which is amongst the deepest UV data ever obtained, to derive the rest frame UV luminosity function at $z\sim 1.7,2.2,3$, and constrain its faint--end slope. In addition, we  use  \ha\ luminosity functions available from the literature to compare the volume averaged SFR derived with the two SFR indicators at $z\sim2$. This paper is organized as follows: Section~\ref{sec:data+sample} describes the UVUDF data used for this work as well as our sample selection; Section~\ref{sec:comp} describes the completeness simulations and presents the selection functions; Section~\ref{sec:LF} outlines the LF fitting procedure as well as our results; Section~\ref{sec:results} presents our results; Section~\ref{sec:discuss} discusses our results in context with recent literature and analyzes the implications; and Section~\ref{sec:conclusions} summarizes our conclusions. Throughout this paper, we assume cosmological parameters from Table 3 of \citet{planck15}: $\Omega_m=0.315$, $\Omega_\lambda = 0.685$ and $H_0=67.31$ km s$^{-1}$ Mpc$^{-1}$ and all magnitudes used are AB magnitudes \citep{oke83}.

\section{Data and Sample Selection}
\label{sec:data+sample}
\subsection{UVUDF Data}
\label{sec:data}
The full UVUDF dataset is comprised of eleven photometric broadband filters covering the \textit{Hubble} UDF ($\alpha(J2000) = 0.3^h32^m39^s, \delta(J2000) = -27\arcdeg47'39."1$) spanning wavelengths from the NUV to NIR. The NUV coverage of the HUDF provided by the UVUDF observations includes three WFC3-UVIS filters: F225W, F275W, F336W \citep{teplitz13}. The optical wavelengths are covered by four ACS filters: F435W, F606W, F775W, F850LP \citep{beckwith06}. The NIR is covered by four WFC3-IR filters: F105W, F125W, F140W, F160W obtained as part of the UDF09 and UDF12 programs \citep{oesch10, bouwens11, koekemoer13, ellis13}. Moreover, the entire field is also covered in F105W, F125W, and F160W as part of the CANDELS GOODS-S observations \citep{grogin11, koekemoer11}. The UVUDF field with coverage in all eleven filters covers an area of 7.3 arcmin$^2$. The data reduction, photometry and source catalog generation for the UVUDF is fully developed and described in \citet{rafelski15}. We use the final catalog provided by \citet{rafelski15} for this work. 

Before applying the sample selection cuts, we remove all sources that are flagged as stars in the UVUDF catalog. Furthermore, we flag bright, compact sources ($z_{850} < 25.5$ and half-light radii, $r_{1/2} < 1"$) with a \texttt{SExtractor} stellarity parameter $>0.8$ as stars. This criterion is only reliable for bright sources and hence, we instead use a color-color cut based on the \citet{pickles98} stellar library at fainter magnitudes to flag stars. At $z_{850} > 25.5$, we flag compact sources ($r_{1/2} < 1"$) that have $V-i$ vs. $i-z$ colors consistent with the \citet{pickles98} stellar sequence to within 0.15 mag as stars. Lastly, we confirm that no stars are left in the final samples by visual inspection.

\subsection{Dropouts Sample Selection}
\label{sec:dropouts}

\begin{figure*}
\centering
\includegraphics[width=0.33\textwidth]{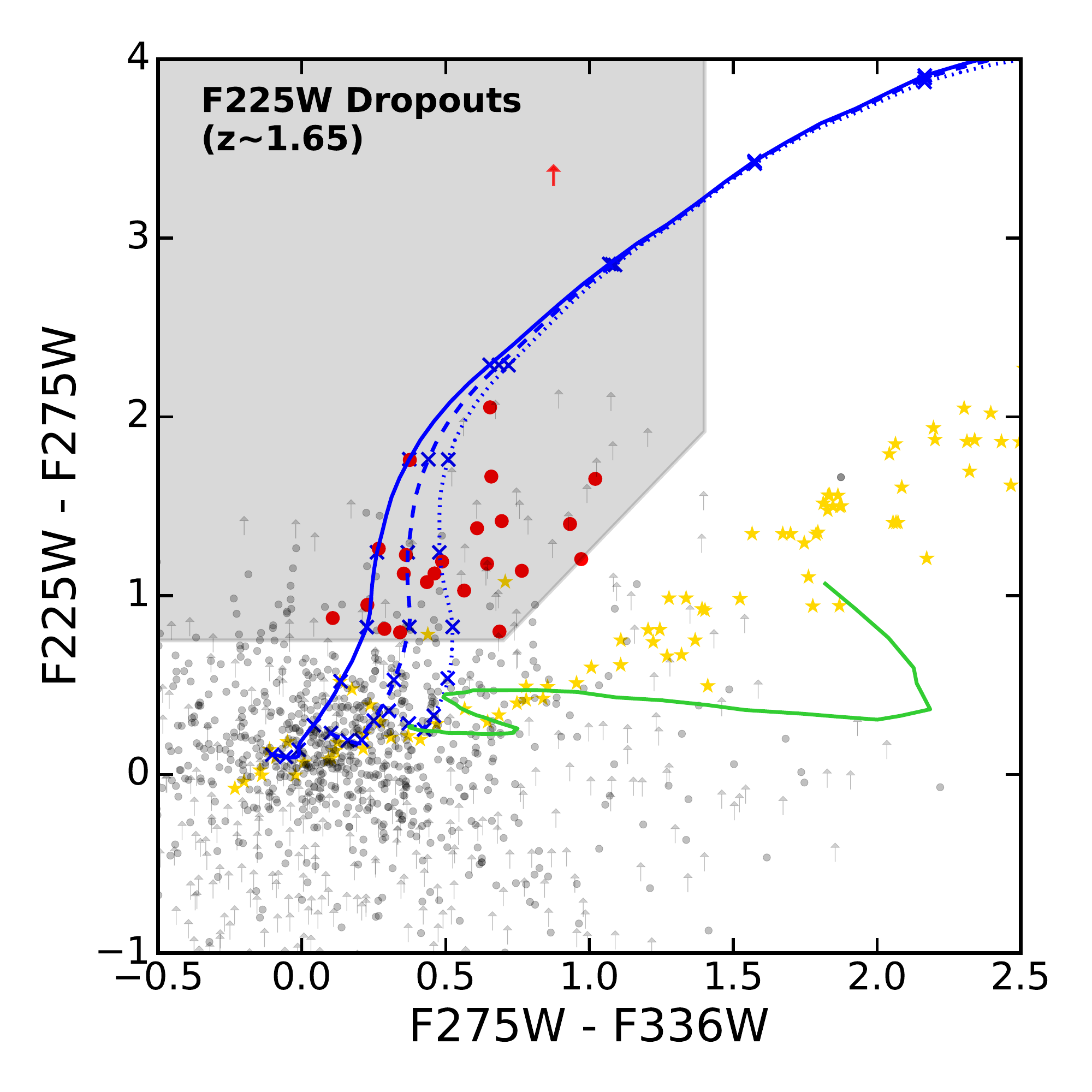}
\includegraphics[width=0.33\textwidth]{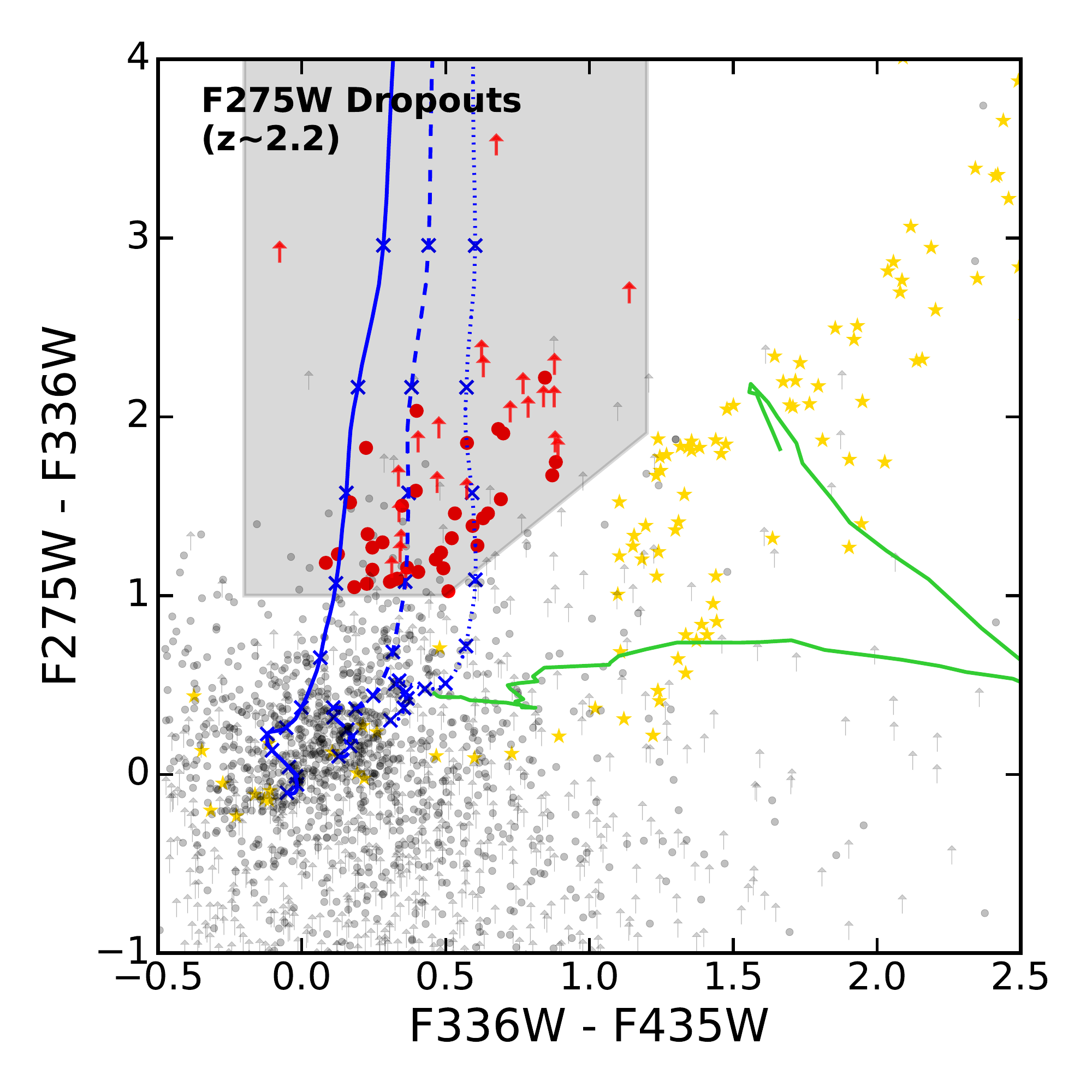}
\includegraphics[width=0.33\textwidth]{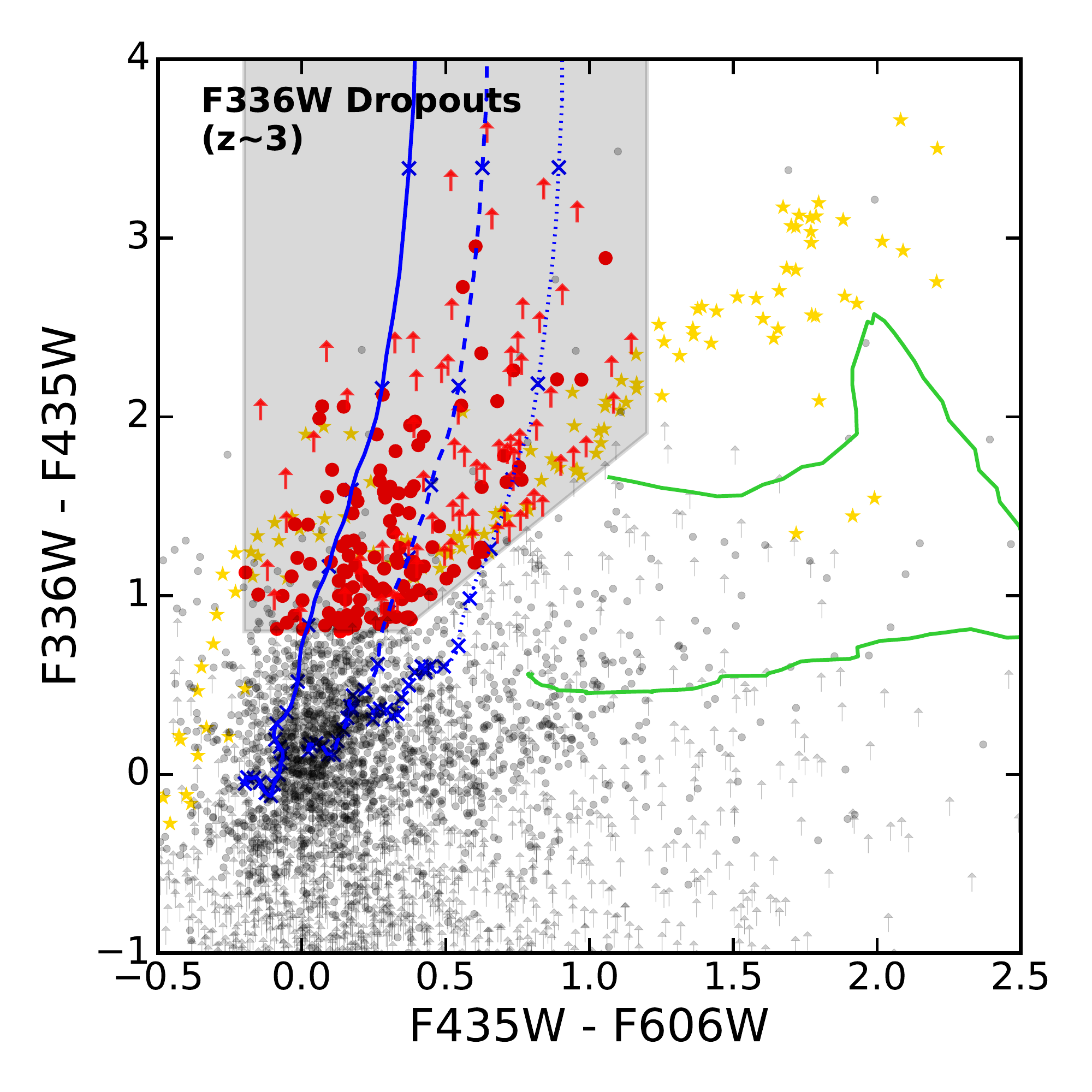}
\caption{The color selection criteria for F225W, F275W, F336W LBG dropouts at $z\sim1.7,2.2,3$ respectively (from left to right). The shaded regions highlight the selection region in color-color space for the dropouts. The \textit{black} points are all detected sources in UVUDF, while the \textit{red} are the ones that make the selection cut. Note the objects selected by the dropout criteria are not only require to be in the shaded region, but also need to make the S/N cuts from Equations~\ref{eqn:dropout225}--\ref{eqn:dropout336}. All sources with fluxes below the 1$\sigma$ limit in the dropout filter have been replaced with their corresponding 1$\sigma$ upper limit. The orange points are stars from \citet{pickles98}, the green lines show the color tracks for low redshift ($0<z<1$) elliptical galaxies from \citet{coleman80}, and the blue lines show color tracks for star-forming galaxies with different dust content, $E(B-V)$ = 0 (\textit{solid}), 0.15 (\textit{dashed}), 0.3 (\textit{dotted}). The star-forming tracks are derived using \citet{bc03} template for constant star-formation rate, solar metallicity, age of 100 Myr and dust extinction defined by \citet{calzetti00} law.}
\label{fig:sample}
\end{figure*}

The Lyman break feature in galaxy SEDs has been proven to be very efficient at selecting high-redshift galaxies \citep[e.g.,][]{steidel96, steidel99, steidel03, adelberger04, bouwens04, bouwens06, bouwens10, bouwens11, bunker04, bunker10, rafelski09, reddy09, reddy12, oesch10, hathi12}. Here, we use the NUV filters available in the UVUDF to identify the Lyman break galaxies (LBGs) in the redshift range of $z\sim1.5-3.5$.

The LBG dropout selection criteria we use, are based on standard color-color and S/N cuts, similar to \citet{hathi10}, \citet{oesch10} and \citet{teplitz13}. However, we further optimize the S/N cuts using the mock galaxy sample generated for our completeness simulations (see Section~\ref{sec:comp}). The color-color selection criteria are shown in Figure~\ref{fig:sample}. Specifically, we select galaxies between $z \sim 1.4 - 1.9$ as follows:

\begin{equation}
\label{eqn:dropout225}
\begin{cases}
F225W - F275W > 0.75 \\
F275W - F336W > -0.5 \\
F275W - F336W < 1.4 \\
F225W - F275W > [1.67 \times (F275W - F336W)] -0.42 \\
F336W - F435W > -0.5 \\
S/N(F275W) > 5 \\
\end{cases}
\end{equation}

\noindent This results in a sample of $25$ galaxies from the UVUDF catalog. Similarly, galaxies between $z \sim 1.8 - 2.6$ are selected using the following criteria:

\begin{equation}
\label{eqn:dropout275}
\begin{cases}
F275W - F336W > 1.0 \\
F336W - F435W > -0.2 \\
F336W - F435W < 1.2 \\
F275W - F336W > [1.3 \times (F336W - F435W)] +0.35 \\
S/N(F336W) > 5 \\
S/N(F225W) < 1
\end{cases}
\end{equation}

\noindent  providing a sample of $60$ galaxies. The galaxies between $z \sim 2.4 - 3.6$ are selected as:

\begin{equation}
\label{eqn:dropout336}
\begin{cases}
F336W - F435W > 0.8 \\
F435W - F606W > -0.2 \\
F435W - F606W < 1.2 \\
F336W - F435W > [1.3 \times (F435W - F606W)] + 0.35 \\
S/N(F435W) > 5 \\
S/N(F275W) < 1
\end{cases}
\end{equation}
which returns $228$ galaxies. When applying these color selection criteria, all sources with magnitudes below the $1\sigma$ limit in the dropout filter are replaced with their corresponding $1\sigma$ upper limits, as determined from our completeness simulations (see Section~\ref{sec:comp}).

\subsection{Photometric Redshift Sample Selection}
\label{sec:photoz}
The inclusion of NUV data (in addition to the optical and NIR) enhances the photometric redshift accuracy, particularly at $z<0.5$ and $2<z<4$ \citep{rafelski15}. The UVUDF catalog includes photometric redshifts calculated using the eleven broadband photometry via Bayesian Photometric Redshift (BPZ) algorithm \citep{benitez00,benitez04,coe06}. The SED templates used for BPZ are based on those from PEGASE \citep{fioc97} recalibrated using redshift information from FIREWORKS \citep{wuyts08}. The quality of the photometric redshift is reported by two quantities, \texttt{ODDS} (measuring the spread in the probability distribution function, $P(z)$) and modified reduced $\chi^2$ (measuring the goodness of fit)\footnote{The modified reduced $\chi^2$ reported by BPZ is similar to a normal reduced $\chi^2$, except it includes an additional uncertainty for the SED templates in addition to the uncertainty in the galaxy photometry \cite{coe06}. The resultant $\chi^2$ is a more realistic measure of the quality of the fit (for more discussion, see \citet{rafelski09}.}. We require the photometric redshift sample to have \texttt{ODDS}$>0.9$ and modified reduced $\chi^2<1$ to ensure selecting only sources with reliable photometric redshifts. Applying these cuts gives a sample of 234, 258 and 440 galaxies in the redshift ranges $1.4<z<1.9$, $1.8<z<2.6$ and $2.4<z<3.6$, respectively.

Using the photometric redshifts enables sample selection down to fainter magnitudes than the corresponding dropout criteria. The dropout selected samples require $5\sigma$ in the detection band to confirm the strength of the break. On the other hand, photometric redshift selected samples only require a $5\sigma$ detection in the rest-1500\AA\ filter. At these redshifts, the dropout detection band (F275W for F225W dropouts, F336W for F275W dropouts, F435W for F336W dropouts) is not the same as the rest-1500\AA\ filter (F435W for $z<2.2$; F606W for $z>2.2$). This is because the dropout detection band looks for flux immediately redward of the Ly$\alpha$ (1216\AA), whereas the rest-frame UV flux is still redward at 1500\AA. Since the optical data available are deeper than NUV, the photometric redshift samples select galaxies down to fainter rest-1500\AA\ magnitudes.

We fit luminosity functions for both the LBG samples as well as photometric redshift-selected galaxy samples in the same redshift ranges as the dropout criteria to validate the fit robustness. Moreover, the depth of the photometric redshift sample allows for better constraints on the faint-end slope.

\section{Completeness}
\label{sec:comp}
Survey incompleteness and sample selection effects greatly impact the effective surveyed volume of a sample, a quantity critical to computing luminosity functions. We need a precise estimate of the completeness for the UVUDF samples in order to properly and accurately correct the volume density. A common approach for completeness estimation in field galaxy studies \citep[e.g.,][]{oesch10,finkelstein15} is to insert mock galaxies into real data, apply identical data reduction and sample selection, and analyze the fraction of recovered artificial galaxies as a function of galaxy properties such as magnitude, redshift, and galaxy size. We perform an extensive set of completeness simulations following a similar procedure in order to quantify the completeness for the UVUDF.

\subsection{Completeness Simulations}
We start by generating a set of mock galaxies with properties representative of the observed sample. These mock galaxies are then planted directly into the real science images, thus preserving the noise properties. Only 150 mock galaxies are inserted at a time to also preserve the crowding properties of the data. These images with artificial galaxies are then put through the same data reduction, analysis for source detection, photometry, photometric redshift, and sample selection as was performed for the real data. By keeping track of the fraction of recovered and selected mock sources compared to the total number of input sources, we can quantify the completeness. Our full set of simulations consists of repeating this process for a total of 45,000 mock galaxies over 300 separate iterations.

To ensure that the mock galaxies used for our completeness simulations are consistent with the observed sample, we assign the absolute magnitudes for our mock galaxies according to existing prescriptions of the UV LFs from the literature. In particular, we use the rest-frame UV LFs from \citet{oesch10} to randomly generate a set of rest-1500\AA\ absolute magnitudes for our mock sample. The initial redshift distribution for our simulated sources is taken to be flat, i.e., $dN/dz$ is constant.

The colors for our mock sample are assigned using spectral templates from \citet[][BC2003]{bc03} models. Each mock galaxy is given a set of model parameters: metallicity, age, exponential SFR $\tau$, and dust extinction. The metallicity is chosen to be random from $Z/Z_\odot$ = 0.0001, 0.0004, 0.004, 0.008, 0.02 (preset BC2003 models). We use the distributions of age, exponential SFR $\tau$, and dust extinction ($E(B-V)$) from observed galaxies in 3D-HST \citep{skelton14} to randomly generate these parameters for our mock galaxies. We also include the contribution from nebular emission lines using line ratios from \citet{anders03}. We apply a \citet{calzetti00} dust extinction law as well as \citet{inoue14} IGM attenuation model to the SEDs to simulate the dust extinction and IGM absorption, respectively. After translating the SEDs to the appropriate redshift, the magnitudes for the rest of the filters are then obtained by computing the contribution of the SED in the particular filter according to its response curve.

We generate the mock galaxies for our completeness simulations using the \texttt{IRAF} task, \texttt{mkobjects}. The sizes for the mock galaxies are defined by assigning a half-light radius for each source. We use the observed distribution of B$_{435}$-band sizes for all sources (no cuts applied) in the UVUDF to randomize the half-light radii for our mock galaxies. The distribution of simulated half-light radii is roughly representative of a log-normal with a peak at 2.7-pixel, with a tail towards larger radii giving an interquantile range of $2.5-4.9$ pixel corresponding to a physical size of $\sim0.6-1.3$ kpc at $z\sim2$. The observed UVUDF catalog shows no significant size bias (in pixels) as a function of redshift for $z<4$ and hence, we choose the size distribution of our mock galaxies to be uniform at all redshifts, in order to fully explore the parameter space. We choose B$_{435}$--band since it is the closest filter in wavelength and has a similar resolution to the rest-frame UV, in addition to the deep coverage available as well as the relatively narrow point-spread function (PSF).

To fully generate an artificial galaxy, \texttt{mkobjects} also requires a Sersic index ($n$), axial ratio, image position, and position angle. The mock galaxies are assigned a Sersic profile that either represents exponential disks ($n=1$; good description of spiral galaxies) or de Vaucoleuers profile ($n=4$; good description of elliptical galaxies). These represent the two extremes for light profiles for observed galaxies. We fix the probability for a mock galaxy to have $n$=1 or $n$=4 to be equal (50\% each). While observed galaxies do not exhibit this distribution, completeness as a function of Sersic index is expected to be well-behaved between the two extremes. Our choice is motivated by wanting to properly sample the two extremes. Furthermore, we verify the simulation output and confirm that the choice of this Sersic index distribution does not bias the completeness in any statistically significant manner.

The ellipticities (axial ratios) for our mock sample are randomized using the distribution of observed B$_{435}$--band axial ratios in the UVUDF, with a peak at 0.7 and long tail towards lower axial ratios. The position of the simulated sources is randomized within a 4000$\times$4400 pixel ($2' \times 2.5'$) region that ensures UV coverage of the UDF. We further limit the positions of mock galaxies to avoid chip edges as well as the WFC3/UVIS chip gap, as is done for the real sample. By allowing for random positions, we can encapsulate any variations in the depth or noise properties across the imaging data. Lastly, the position angles are randomized between 0\degr and 360\degr.

We determine the optimal number of galaxies to insert in one iteration in order to avoid crowding issues by planting a varying number of mock galaxies into the images. From this, we find that the scatter in the recovered photometry compared to the input does not increase significantly between 100-200 sources inserted per iteration. Thus, planting $\lesssim200$ sources per iteration does not change the crowding properties of our field. Being conservative, we choose to insert 150 sources per iteration. This is similar to the treatment for completeness simulations done for CANDELS/GOODS fields by \citet{finkelstein15}.

We split the simulation into 100 separate iterations and, for each iteration, we insert 150 mock galaxies into the original images. The mock galaxies are simulated and inserted into the original image for each of the eleven filters using \texttt{mkobjects}. The newly generated images with simulated sources are then run through identical data reduction, photometry and catalog creation process as for the real data.

Briefly, the process involves using \texttt{ColorPro} \citep{coe06} to measure photometry in the images, which runs Source Extractor \citep{bertin96} for all filters in dual-image mode. \texttt{ColorPro} also applies aperture corrections for the different aperture sizes as well as PSF corrections to account for variations in the PSF across filters. The detection image used is created from the 4 optical and 4 NIR filters to maximize depth and robustness of the aperture sizes. In order to properly recover both bright and faint sources in a crowded field, photometry is measured using two different detection thresholds and two different deblending thresholds, which is then merged into a single photometric catalog. For the smaller apertures needed for the NUV filters, F435W is used as the detection image, instead. The full methodology is described in detail in \citet{rafelski15}.

The UVUDF catalog includes aperture-matched PSF corrected photometry for a robust measurement of flux across images with varying PSFs. This is done by measuring photometry on high-resolution data and applying a PSF correction for the NIR filters, which have larger PSFs. The PSF correction is determined by degrading the I$_{775}$ image (reddest high-resolution image with a well-behaved PSF available) to each of the NIR filters using the \texttt{IRAF} task, \texttt{psfmatch}. Instead of convolving the entire image after mock galaxies are inserted (which is computationally expensive), we instead create 500$\times$500px (15\arcsec$\times$15\arcsec) stamps for the simulated sources using \texttt{mkobjects} and convolve them individually before adding to the PSF matched image. This saves considerable amount of computation time per iteration.

Lastly, we generate a catalog using the images with mock galaxies following the same pipeline as used for the real UVUDF catalog \citep{rafelski15} and compare it to the input catalog to determine the fraction of recovered objects. In order to be recovered, an object is required to be positionally matched within 3px (0.09\arcsec). 

Typically, incompleteness is computed as a function of magnitude and redshift; however, we also consider another key factor: galaxy size. Even at a constant magnitude and redshift, an extended galaxy may not be recovered due to the low surface brightness compared to a compact one. We correct the effective volumes for our sample according to the magnitude, redshift as well as the galaxy size using selection functions as described in Section~\ref{sec:self}. We also compute the completeness as a function of apparent magnitude and galaxy size, which is used to define the survey magnitude limits (see Appendix~\ref{appndx:comp}).

\begin{figure*}[h]
\centering
\includegraphics[width=0.33\textwidth]{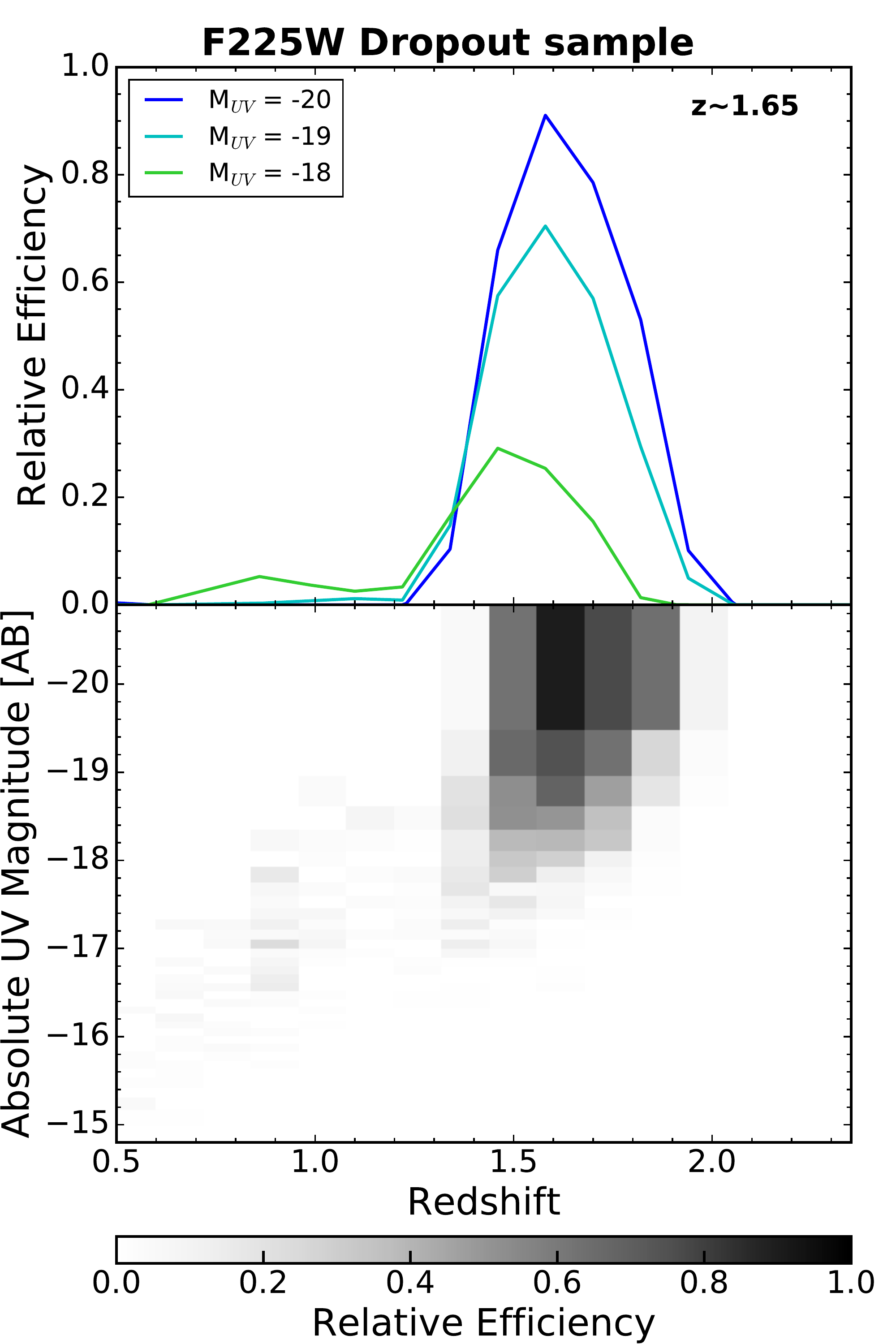}
\includegraphics[width=0.33\textwidth]{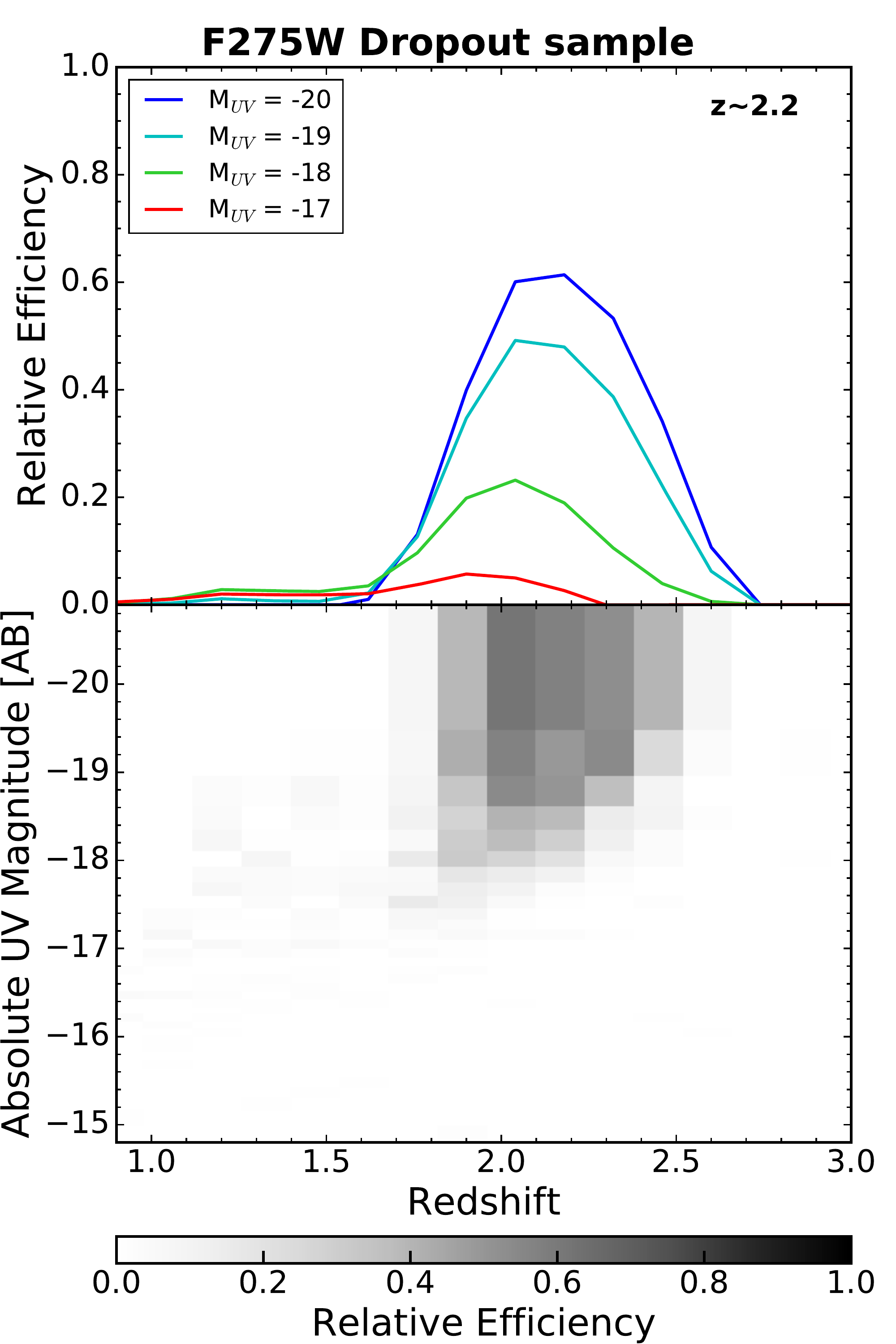}
\includegraphics[width=0.33\textwidth]{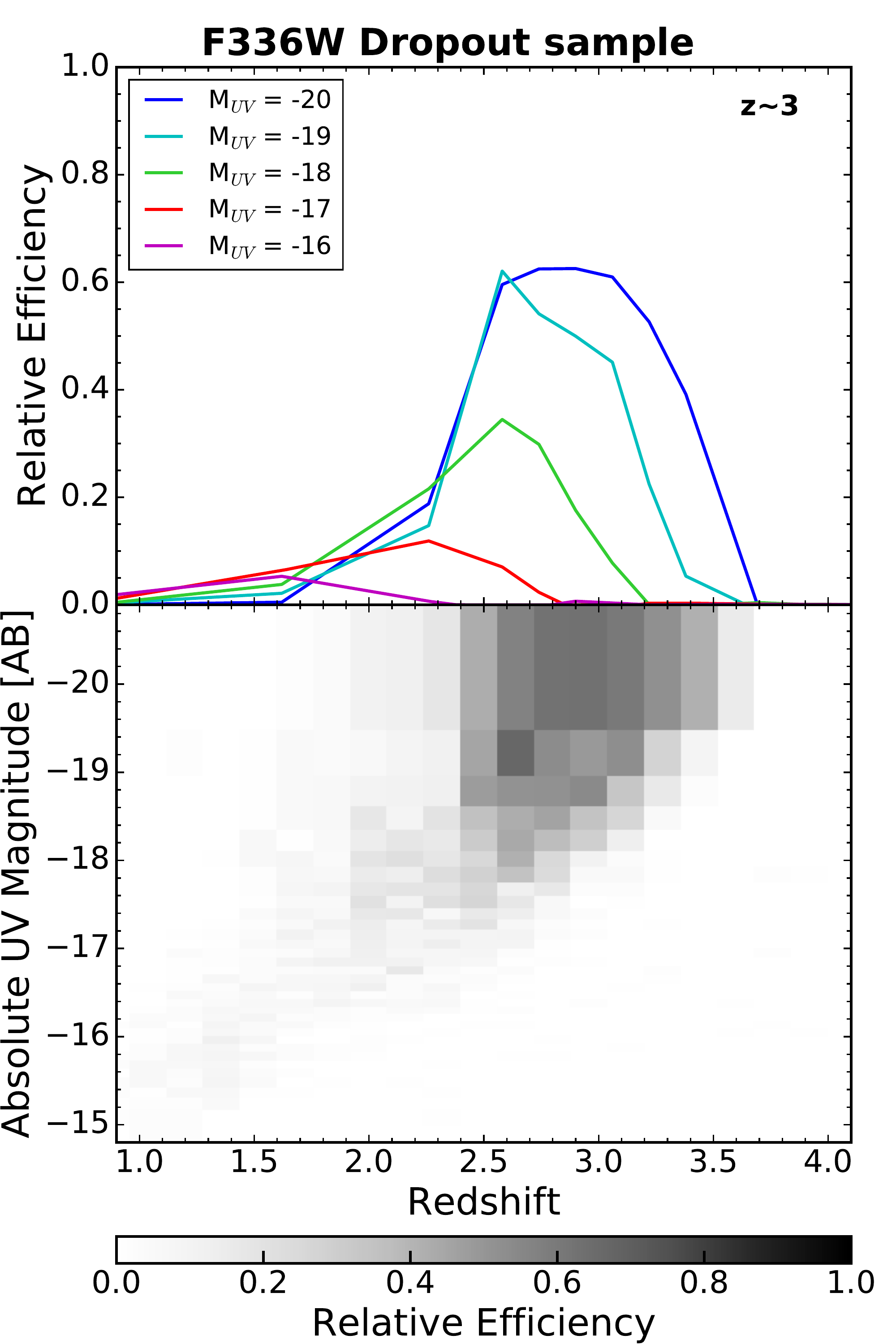} \\
\vspace{0.2in}
\includegraphics[width=0.33\textwidth]{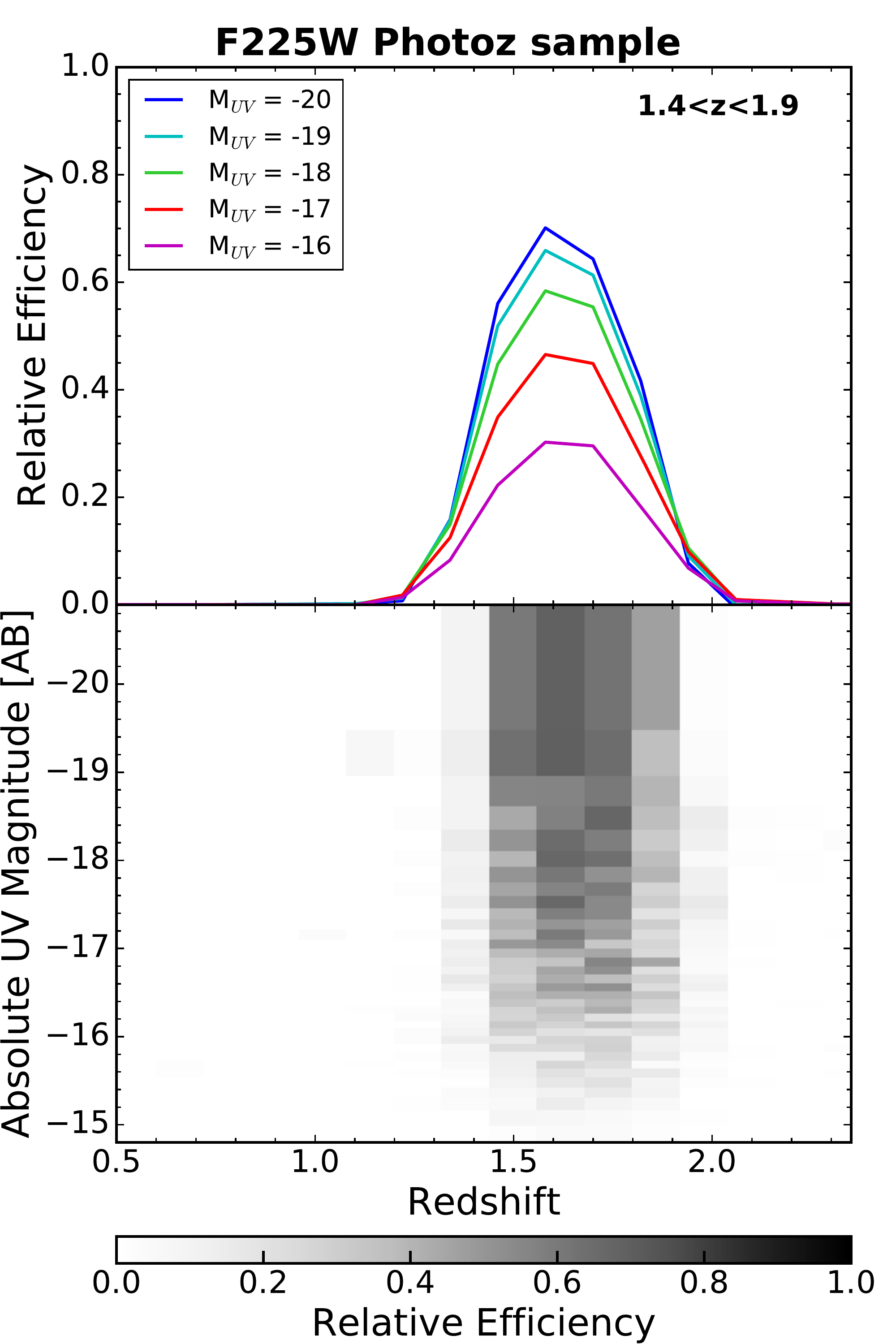}
\includegraphics[width=0.33\textwidth]{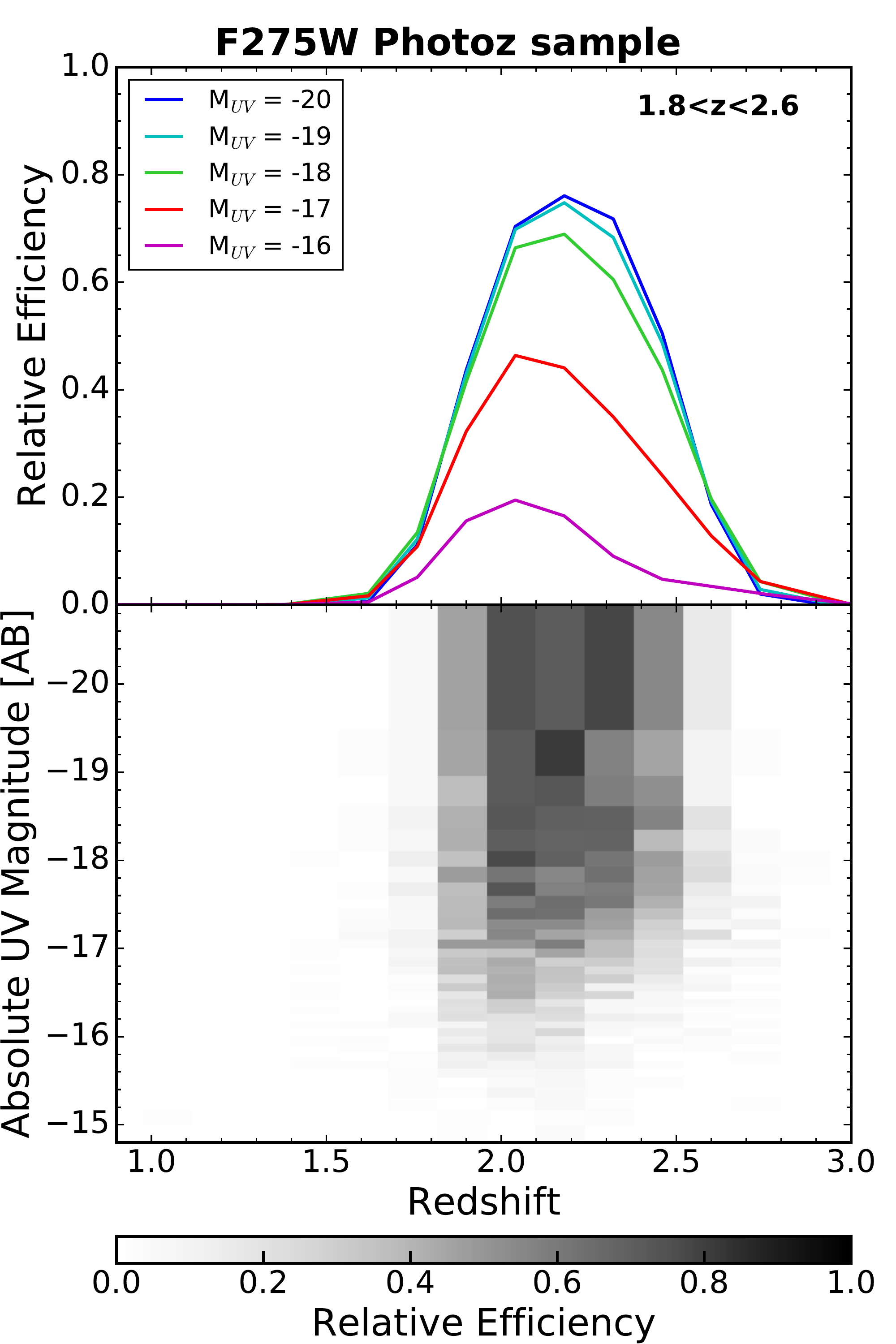}
\includegraphics[width=0.33\textwidth]{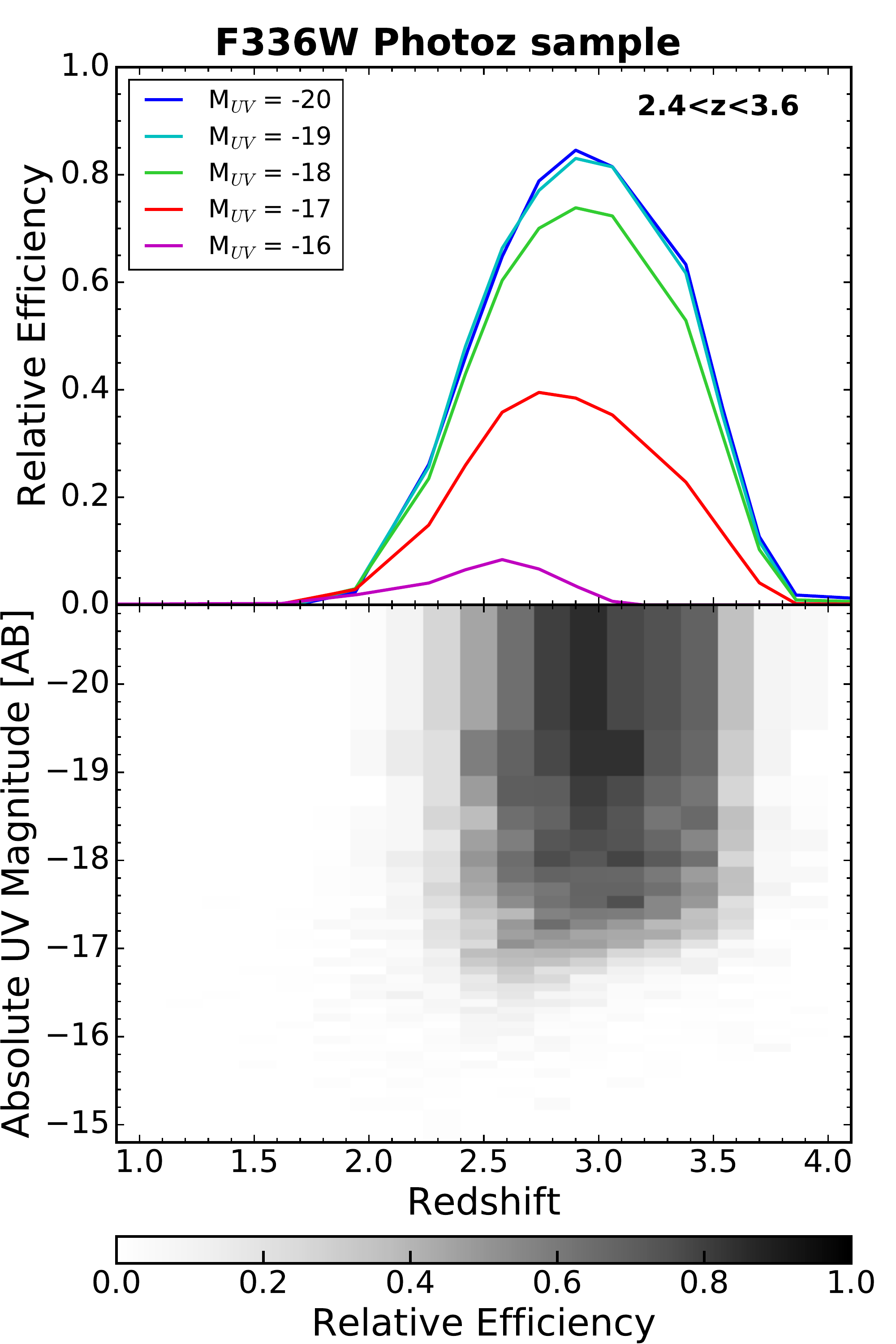}
\caption{(\textit{Top row}) Selection functions for the F225W, F275W and F336W dropout criteria shown as a function of redshift (top panel) and as a function of both redshift and absolute rest-frame UV magnitude (bottom panel). This shows the redshift and absolute rest-frame UV magnitude distributions of galaxies selected by the dropout criteria. The relative efficiency is the fraction of sources that are selected from the full input sample after applying the selection cuts. (\textit{Bottom row}) Selection functions for the photometric redshift samples corresponding to the redshift selected by dropout criteria. Requiring a 5$\sigma$ detection in the observed rest-frame UV magnitude and applying the cuts to ensure good quality photometric redshifts (\texttt{ODDS}$>0.9$ and modified reduced $\chi^2<1.0$) are the primary factors affecting the relative efficiency for these selection functions.}
\label{fig:self}
\end{figure*}

\subsection{Selection Functions}
\label{sec:self}
The mock galaxy sample can be used to determine the probability of a galaxy to satisfy the LBG selection criteria as long as there are no biases in the recovered magnitudes and colors of the mock galaxies. We verify that our completeness simulations do not introduce any offsets or biases in the magnitudes and colors of recovered mock galaxies, before proceeding. We apply the LBG selection criteria to the input mock sample to quantify the relative efficiency of the criteria to select a galaxy with given redshift, rest-1500\AA\ absolute magnitude and half-light radius. The top row in Figure~\ref{fig:self} shows the selection function for the three dropout criteria from Section~\ref{sec:dropouts}, marginalized over the all galaxy sizes.

Similarly, the photometric redshift selection criteria are also applied to the mock catalog and the corresponding selection functions are derived for the photometric redshift samples. These are plotted in the bottom row of Figure~\ref{fig:self}, again marginalized over all galaxy sizes. These selection functions are used to compute the effective volumes when fitting the UV LF in Section~\ref{sec:LF}.

The selection functions plotted in Figure~\ref{fig:self} have been marginalized over all galaxy sizes. However, the information in the size dimension is preserved and the effective volume for each source is computed according to its size. The overall effect of galaxy sizes on the completeness can be visualized by Figure~\ref{fig:comp} in Appendix~\ref{appndx:comp}, which shows the survey incompleteness as a function of observed magnitude and galaxy size.

\subsection{Redshift distribution of the Dropout Sample}
\begin{figure*}
\centering
\includegraphics[width=\textwidth]{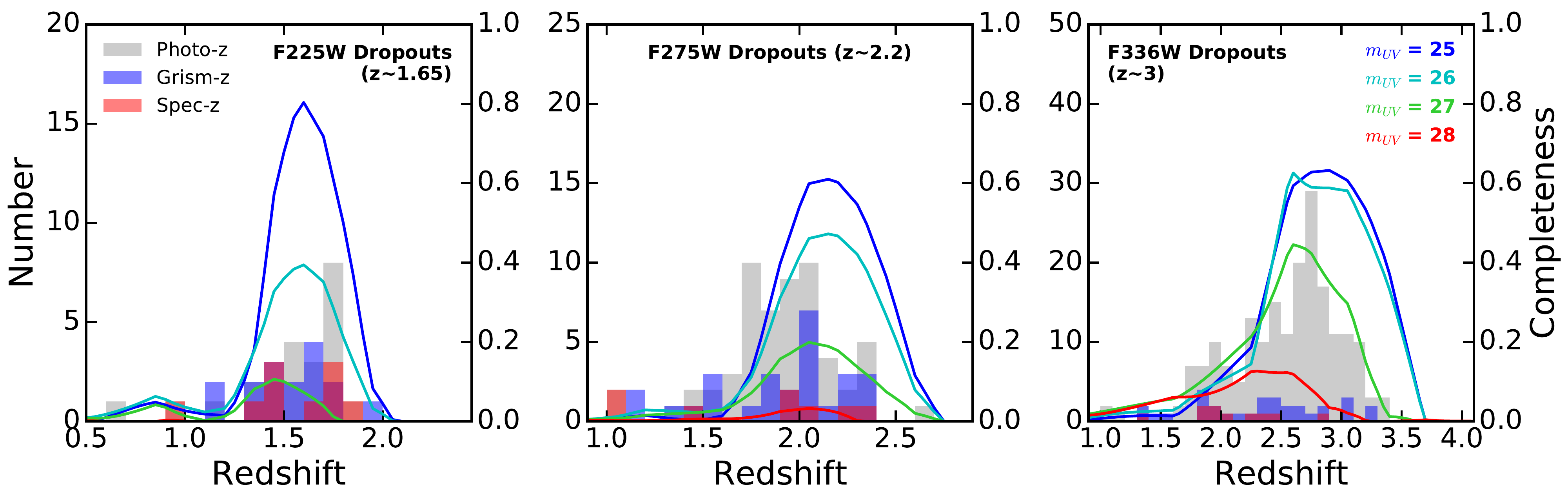}
\caption{The redshift distribution of the dropout-selected samples. The \textit{grey}, \textit{blue} and \textit{red} histograms show the distribution of photometric, grism, and spectroscopic redshifts for the dropout samples, respectively. The curves show the completeness associated with the corresponding dropout criteria for a range of observed UV magnitudes, as computed from our completeness simulations (Section~\ref{sec:comp}). The right-hand axis gives the scale for the completeness values.}
\label{fig:check_z_dropout}
\end{figure*}

The selection functions computed here, when marginalized over the magnitude and size dimensions, describe the redshift distribution of galaxies in the selected sample. As a validation check, we compare the selection functions from our simulations with redshifts measured from observations. Spectroscopic redshifts are available in the UVUDF catalog \citep{rafelski15} for a small fraction of the sources in our sample, whereas photometric redshifts are available for all the sources in the UVUDF catalog. Figure~\ref{fig:check_z_dropout} shows the distribution of spectroscopic (where available) as well as photometric (for all sources) redshifts. The selection functions derived from our completeness simulations are over-plotted for comparison. As seen in the figure, the redshift distributions (shown as histograms) are in overall agreement with the selection functions (shown as curves). Note that the photometric redshifts and simulations are not expected to agree one-to-one by construction, because the galaxy templates used to estimate the photometric redshifts (Section~\ref{sec:photoz}) are not the same as those used to generate the mock catalog for our simulations (Section~\ref{sec:comp}).

\section{Deriving the Luminosity Function parameters}
\label{sec:LF}
The rest-frame 1500\AA\ UV LF is one of the key diagnostics for establishing the link between galaxy luminosities, galaxy masses, and the cosmic star formation rate. Considerable effort has been put into characterizing the shape and evolution of the UV LF near the peak of cosmic star formation ($z \sim 1 - 3$) \citep[e.g.,][]{reddy09,hathi10,oesch10,sawicki12,parsa16,alavi14,alavi16}. Here, we fit UV LFs using the dropout as well as photometric redshift selected samples described in Section~\ref{sec:data+sample} corresponding to the three redshift ranges: $z \sim 1.7,2.2,3.0$.

Using the Schechter function \citep{schechter76} as the parametric shape for the UV LF is well motivated, as it matches the observed Universe well:

\begin{equation}
\label{eqn:sch}
\phi( M) = 
0.4 \ln{(10)} \ \phi^\star
10^{-0.4(M-M^\star)(1+\alpha)}
e^{- 10^{-0.4(M-M^\star)}}
\end{equation}

We perform a maximum likelihood analysis to fit the UV LFs. Specifically, we use the modified maximum likelihood estimator (MLE) developed and presented in \citet{mehta15}, which accounts for the measurement errors in galaxies' observed magnitude, allowing for a more robust fitting procedure. Following the procedure from \citet{mehta15}, we define the probability for detecting a galaxy in the sample as:

\begin{equation} 
\label{eqn:prob}
P(M_i) =
\frac{
\begin{split}
\int \int _{-\infty} ^{M_{lim}(z)}
\phi(M) \cdot & v_{\mathrm{eff}}(M,z) \cdot \\
&  N(M|\{M_i, \sigma_i\}) \mathrm{d}M\mathrm{d}z
\end{split}
}
{\displaystyle \int \int _{-\infty} ^{M_{lim}(z)} \phi(M) \cdot v_{\mathrm{eff}}(M,z) \cdot \mathrm{d}M\mathrm{d}z} \\
\end{equation}

Here, $M_i$ is the absolute rest-1500\AA\ (UV) magnitude of the galaxy, $\phi(M)$ is the luminosity function from Equation~\ref{eqn:sch}, $M_{lim}(z)$ represents the survey's detection limit, $v_{\mathrm{eff}}(M,z)$ is the effective differential comoving volume and $N(M|\{M_i,\sigma_i\})$ is the term that marginalizes over the measurement error in the galaxy's magnitude. The effective differential comoving volume is defined as:

\begin{equation}
\label{eqn:veff}
v_{\mathrm{eff}}(M,z) =  \frac{dV_{comov}}{dz \ d\Omega}(z) \cdot S(M,z,r_{1/2}) \cdot \Omega \\
\end{equation}

\noindent where $S(M,z,r_{1/2})$ is the selection function, which defines the efficiency of the selection criterion at a given redshift $z$ as a function of the absolute UV magnitude $M$ for a given half-light radius $r_{1/2}$ derived from the completeness simulations in Section~\ref{sec:comp}, and $\Omega$ is the solid angle surveyed. The measurement error associated with the absolute UV magnitude ($\sigma_{M,i}$) is modelled as a Gaussian:

\begin{equation}\label{eqn:gauss}
N(M|\{M_i, \sigma_i\}) = \frac{1}{\sqrt{2\pi}\sigma_i} \mathrm{exp} \left[ - \left( \frac{(M - M_i)^{2}}{2 \sigma_{M,i}^{2}}\right)\right]
\end{equation}

In the MLE formalism, the Schechter function normalization $\phi^\star$ is calculated after finding the best-fit values for $\alpha$ and $M^\star$:

\begin{equation}\label{eqn:phi}
\phi^{\star} = \frac{N}{ \displaystyle \int \int _{-\infty}^{M_{lim}(z)} \phi(M) \cdot v_{\mathrm{eff}}(M,z) \cdot \mathrm{d}M\mathrm{d}z}
\end{equation}

We construct the log likelihood function, $\ln{\mathcal{L}} = \sum_{i=1}^N \ln{P(M_i)}$, where the probability for each source to be detected in the sample, $P(M_i)$ is computed using Equation~\ref{eqn:prob}. We maximize the log likelihood function (alternatively, minimize the negative log likelihood function). Once the best fit values for slope $\alpha$ and characteristic magnitude $M^\star$ are obtained, the normalization $\phi^\star$ is calculated.

Furthermore, in order to properly quantify the uncertainties on our best-fit parameters, we perform a Markov Chain Monte Carlo analysis (MCMC). We probe the full posterior distribution for the free parameters in LF fitting using the Python package \texttt{emcee} \citep{foreman13}. We implement the Affine-Invariant Ensemble Sampler in \texttt{emcee}, initialized at the best-fit parameters. The uncertainties on our LF parameters are obtained from the distribution of the Markov chain, after discarding the burn-in period. 

\section{Results}
\label{sec:results}

\begin{figure*}
\centering
\includegraphics[width=0.7\textwidth]{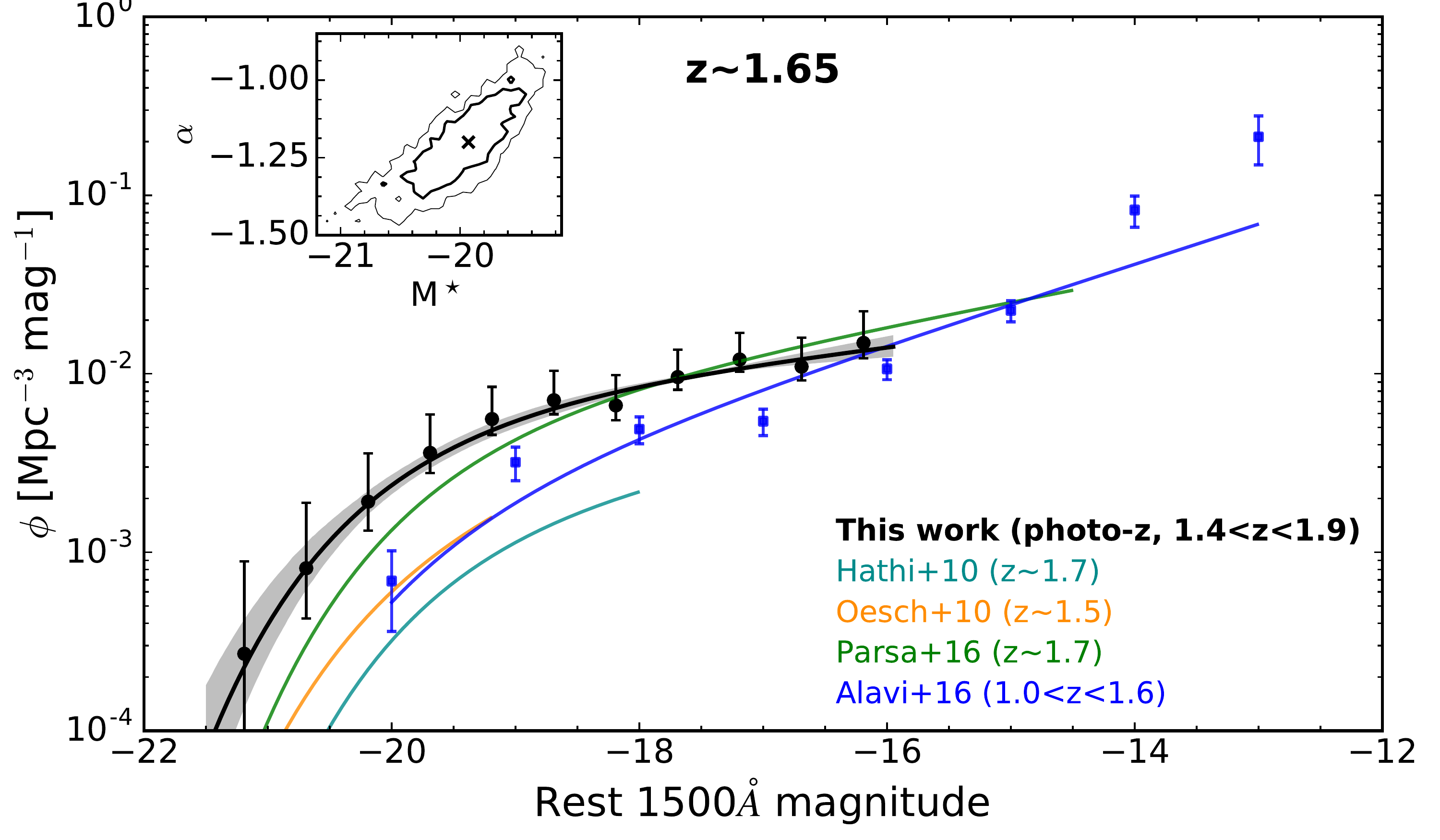}
\includegraphics[width=0.7\textwidth]{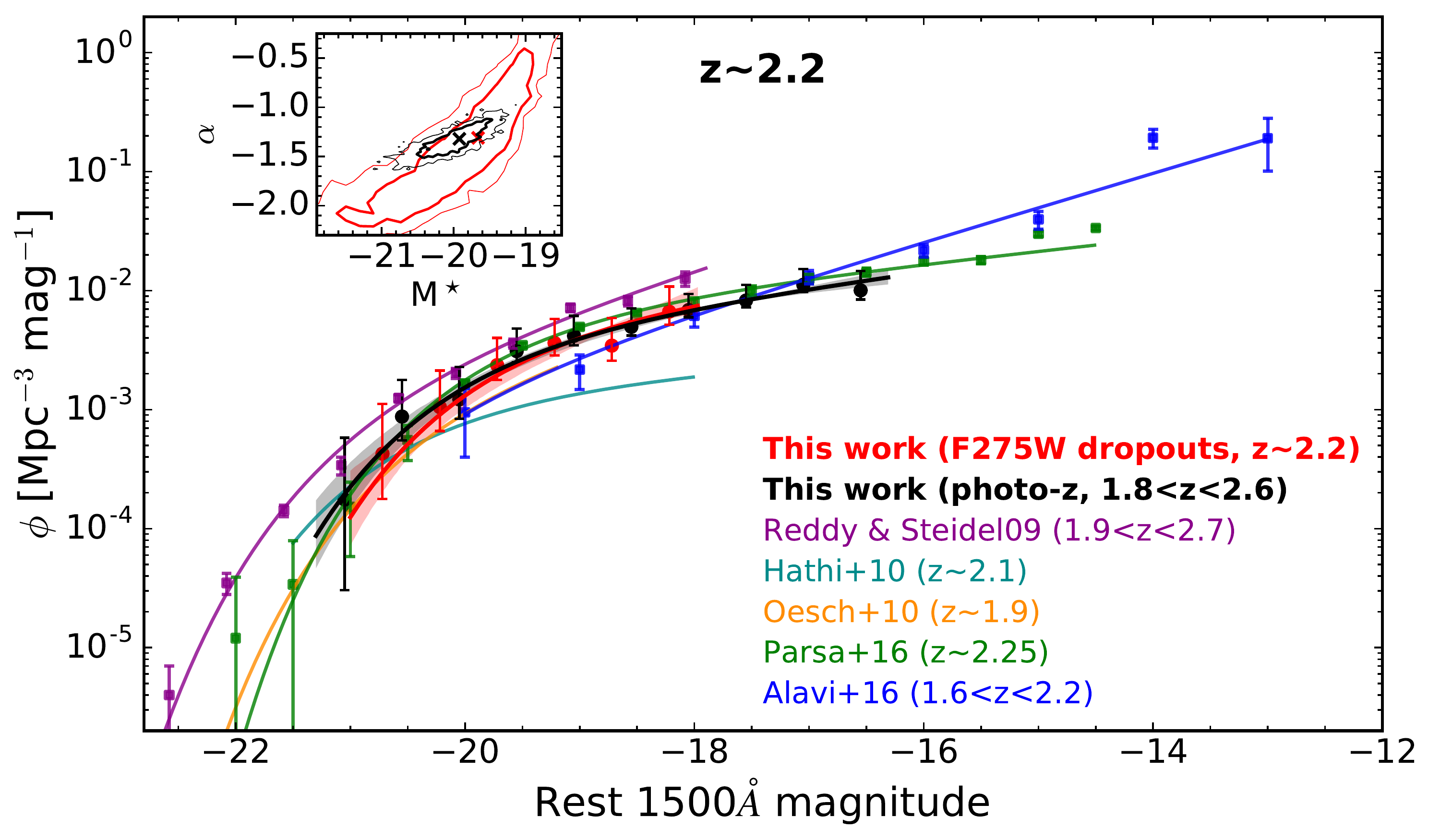}
\includegraphics[width=0.7\textwidth]{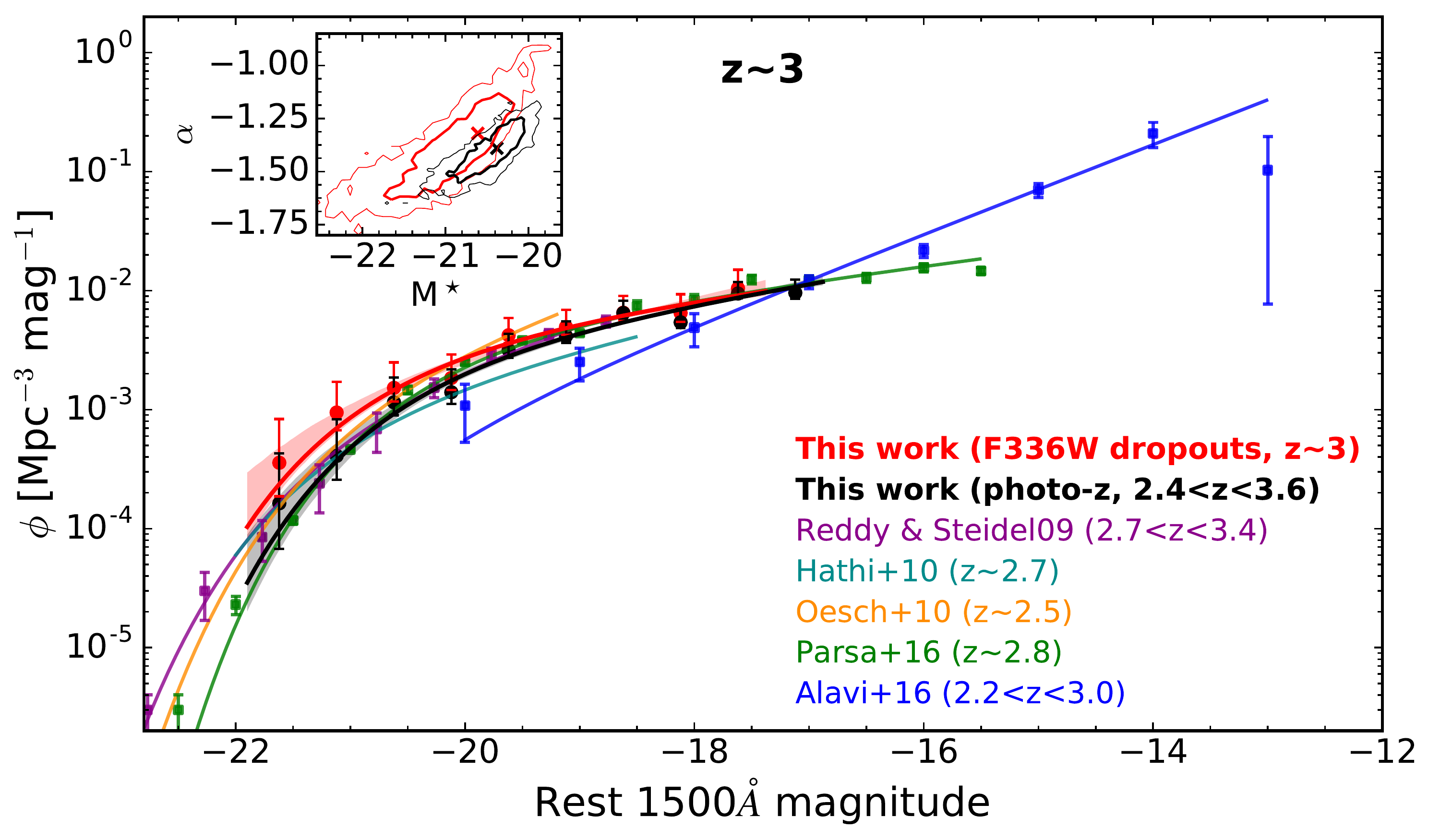}
\caption{The rest-frame 1500\AA\ UV luminosity functions at $z\sim1.7$ (bottom left), $z\sim2.2$ (top), and $z\sim3$ (bottom right) measured using UVUDF. The best-fits obtained using the dropout samples from UVUDF are shown in \textit{red}, and the corresponding photometric redshift samples are shown in \textit{black}. The best-fit parameters for all luminosity function fits are reported in Table~\ref{tab:pars}. The insets show the 68\% (\textit{thick}) and 95\% (\textit{thin}) confidence regions for the free luminosity function parameters ($\alpha$ and $M^\star$) obtained from MCMC analysis. The shaded regions denote the 1$\sigma$ confidence regions for the UV luminosity function fits. We also plot the UV luminosity functions at similar redshifts from recent literature for comparison. All LFs have been plotted for the range of magnitudes covered by their samples.}
\label{fig:LF}
\end{figure*}

\subsection{Rest-frame UV Luminosity functions at $z\sim1.5-3$}
\label{sec:LFfit}
Using the modified MLE fitting procedure described in Section~\ref{sec:LF}, we fit a rest-frame UV LF for the dropout as well as photometric redshift selected samples from Section~\ref{sec:data+sample}. The photometric redshift samples selected over the same redshift ranges as those covered by the dropout samples allows us to verify the robustness of our LF fits.

Ideally, one would use the full sample selected down to the faintest magnitudes possible within the survey capabilities. However, near the survey limit, the incompleteness rises significantly and the correction applied to the effective volumes tends towards considerably large values. In order to avoid using sources with corrections that are too large, we choose to truncate the sample where the effective volume correction rises above 75\% of the correction at the bright end, which further reduces  our sample size. Table~\ref{tab:pars} reports the final sample sizes used to fit the LFs.

\subsubsection{UV LF at $1.4<z<1.9$}
The F225W dropout selection criterion results in a sample of 23 galaxies from the UVUDF catalog. Due to the small sample size, we choose not to fit a LF for the F225W dropouts. The corresponding photometric redshift sample ($1.4<z<1.9$), however, consists of 202 galaxies -- sufficient to properly fit a LF. We use the F435W as the rest-1500\AA\ magnitude and the effective volumes corrected using the selection functions from Section~\ref{sec:comp}. The resulting best-fit parameters are reported in Table~\ref{tab:pars} and the LF is plotted in the bottom-left panel of Figure~\ref{fig:LF} along with the results from recent literature.

The UV LFs available from the literature at this redshift already show a considerable spread in their parameters. Our best-fit UV LF expects a higher number density at the bright end compared to other LFs. This can be inferred from the high $M^\star$ value we find for our best-fit LF. However, it is important to point out that the area covered by the UVUDF survey is small (7.3 arc. min$^2$), which leads to high cosmic variance. The bright end of the LF is particularly prone to this, given the small number statistics. The faint-end slope of our best-fit LF is also considerably shallow compared to other LFs at similar redshifts. It is important to note that a higher $M^\star$ value also contributes towards flattening the faint-end slope. Our $\alpha$ value still agrees with the F225W dropout LFs from \citet{hathi10} and \citet{oesch10} within their uncertainties. There is minimal tension between our best-fit and \citet{parsa16}, as their value is within $<1.5 \sigma$ of ours. However, there is a substantial discrepancy between our result and \citet{alavi16} value of $\alpha=1.56\pm0.04$.

\subsubsection{UV LF at $1.8<z<2.6$}
The F275W dropout criterion selects galaxies with $1.8<z<2.6$, where the rest-1500\AA\ magnitude is covered by F435W for $z<2.2$ and F606W for $z>2.2$. For the photometric redshift sample, we use the appropriate rest-1500\AA\ filter identified using the photometric redshift. However, for the dropout sample, this is not possible due to the lack of individualized redshift information; instead, we use F435W as the rest-1500\AA\ filter for the full sample, since it covers rest-1500\AA\ for the majority of the redshift range (considering the longer tail towards lower redshift). We fit a rest-frame UV LF for the sample of 58 galaxies selected by the F275W dropout criterion, with effective volumes corrected according the corresponding selection function. We also fit a LF using the $1.8<z<2.6$ photometric redshift sample consisting of 238 galaxies. The LF fit using the photometric redshift sample agrees with the dropout sample LF, within the $1\sigma$ uncertainties. Both $z\sim2.2$ rest-frame UV LF fits are plotted in the top panel of Figure~\ref{fig:LF} along with the 68\% confidence regions on the free parameters ($\alpha$ and $M^\star$) in the inset, and the best-fit parameters are reported in Table~\ref{tab:pars}.

The nature of the UVUDF observations highlights the ability to go to faint, albeit in a small area. Hence, one of the main goals of this work is to constrain the faint-end slope of the UV LF. At $z\sim2.2$, we use the UVUDF photometric redshift sample to fit a UV LF faint-end slope of $\alpha=-1.32^{+0.10}_{-0.14}$, which in good agreement with \citet{parsa16} and \citet{sawicki12}, given their uncertainties. On the other hand, our result is considerably shallower than the estimates from \citet{oesch10} and \citet{alavi16}, who find $\alpha=-1.60\pm0.51$ (for their F275W dropout sample) and $\alpha=-1.73\pm0.04$, respectively. However, our sample goes $\sim$3 magnitudes deeper than \citet{oesch10}, thus providing a tighter constraint on the faint-end slope. \citet{alavi16} derive their UV LF using lensed galaxies in the Abell 1689 cluster as well as Abell 2744 and MACSJ0717 clusters from in the Hubble Frontier Fields (HFF). Although, \citet{alavi16} go much deeper (down to $M_{UV}=-13$) than the blank field surveys, there is a possibility of significant systematics affecting their result.

\cite{bouwens16} assess the impact of systematic errors in the fits of the LFs derived from lensed galaxy surveys. They find considerable systematic scatter for faint, high magnification sources ($\mu>20$) dependent on the lens model used, which in turn, has a significant impact on the recovered LF. Most dramatically, they find that the faint-end of the recovered LF is preferentially steeper than the real value, when the systematic uncertainties in $\mu$ are not accounted for. This systematic could help resolve the tension between our result and \citet{alavi16}.

\subsubsection{UV LF at $2.4<z<3.6$}
The F336W dropout sample has 201 galaxies and the corresponding photometric redshift sample consists of 412 galaxies. F606W covers rest-1500\AA\ filter for the redshift range selected by F336W dropouts ($2.4<z<3.6$). Our best fit values for the Schechter parameters along with the uncertainties for all our fits are reported in Table~\ref{tab:pars}. The bottom right panel of Figure~\ref{fig:LF} shows the UV LFs for this redshift range, for both dropout and photometric redshift samples, in comparison with the results from recent literature.

Our best-fit rest-frame UV LF at $z\sim3$ is in excellent agreement with \citet{parsa16}. Our faint-end slope value of $\alpha=-1.39^{+0.08}_{-0.12}$ is considerably shallower than the \citet{reddy09} and \citet{oesch10} value of $\alpha \sim -1.73$. Similar to the F225W and F275W dropouts, there is significant tension when comparing our result to the \citet{alavi16} value of $\alpha \sim -1.94\pm0.06$ fit at a slightly lower redshift $z \sim 2.7$.

\begin{deluxetable*}{ccccccc}
\tablecaption{Best fit parameters for UV LFs}
\tablewidth{0 pt}
\tablehead{ 
\colhead{Redshift} & \colhead{Sample Selection} & \colhead{$M_{lim,\mathrm{UV}}$} & \colhead{$N$\tablenotemark{a}} & \colhead{$\alpha$} & \colhead{$M^\star$} & \colhead{log $\phi^\star$}
}
\startdata

\cutinhead{LBG dropout samples}

$z\sim1.65$ & F225W dropouts & -18.46 & 23 & \multicolumn{3}{c}{sample size too small} \\
$z\sim2.2$ & F275W dropouts & -17.97 & 58 & $-1.31^{+0.32}_{-0.75}$ & $-19.66^{+0.32}_{-1.67}$ & $-2.21^{+0.18}_{-1.13}$ \\
$z\sim3.0$ & F336W dropouts & -17.37 & 201 & $-1.32^{+0.07}_{-0.26}$ & $-20.61^{+0.12}_{-0.92}$ & $-2.36^{+0.12}_{-0.19}$ \\

\cutinhead{Photometric redshift samples}

$1.4<z<1.9$ & Photo-$z$& -15.94 & 202 & $-1.20^{+0.10}_{-0.13}$ & $-19.93^{+0.25}_{-0.40}$ & $-2.12^{+0.12}_{-0.19}$ \\
$1.8<z<2.6$ & Photo-$z$& -16.30 & 238 & $-1.32^{+0.10}_{-0.14}$ & $-19.92^{+0.24}_{-0.44}$ & $-2.30^{+0.12}_{-0.23}$ \\
$2.4<z<3.6$ & Photo-$z$ & -16.87 & 412 & $-1.39^{+0.08}_{-0.12}$ & $-20.38^{+0.19}_{-0.43}$ & $-2.42^{+0.11}_{-0.21}$ \\

\vspace{-0.12in}
\enddata
\tablenotetext{a}{Sample size after removing any sources with high effective volume correction. See Section~\ref{sec:LFfit} for details.}
\label{tab:pars}
\end{deluxetable*}

\subsection{Cosmic Variance}

The errorbars shown in Figure~\ref{fig:LF} already account for the Poisson errors on the number counts. However, given the small field-of-view of the UVUDF, the number counts are also affected by cosmic variance. This would help explain the discrepancy at the bright end of the lowest redshift ($z\sim1.65$) LF compared to other surveys with large coverage. We estimate the cosmic variance for our sample using the Cosmic Variance Calculator v1.02\footnote{\url{http://casa.colorado.edu/~trenti/CosmicVariance.html}} \citep{trenti08}. For a field-of-view of $2.7' \times 2.7'$, we estimate a fractional error of $0.21, 0.21, 0.18$ on the number counts of bright ($M_{UV} < -20$) sources in our $1.4<z<1.9$, $1.8<z<2.6$, $2.4<z<3.6$ photometric redshift samples, respectively. Similarly, the F275W and F336W LBG dropout samples are affected by a fractional error of $0.22, 0.17$ on the number counts of $M_{UV} < -20$ sources, respectively.

\subsection{UV Luminosity Density}
\label{sec:uvld}

The faint-end slope of the UV LF determines the relative contribution of faint and bright galaxies to the total cosmic UV luminosity. We use the new estimates derived in Section~\ref{sec:LFfit} to compute the observed cosmic UV luminosity density (not corrected for dust) as:

\begin{equation}
\label{eqn:uvld}
\rho_{UV} = \int_{L_{lim}}^{\infty} L\phi(L)dL = \int_{-\infty}^{M_{lim}} L(M)\phi(M) dM
\end{equation}

\begin{figure}
\centering
\includegraphics[width=0.45\textwidth]{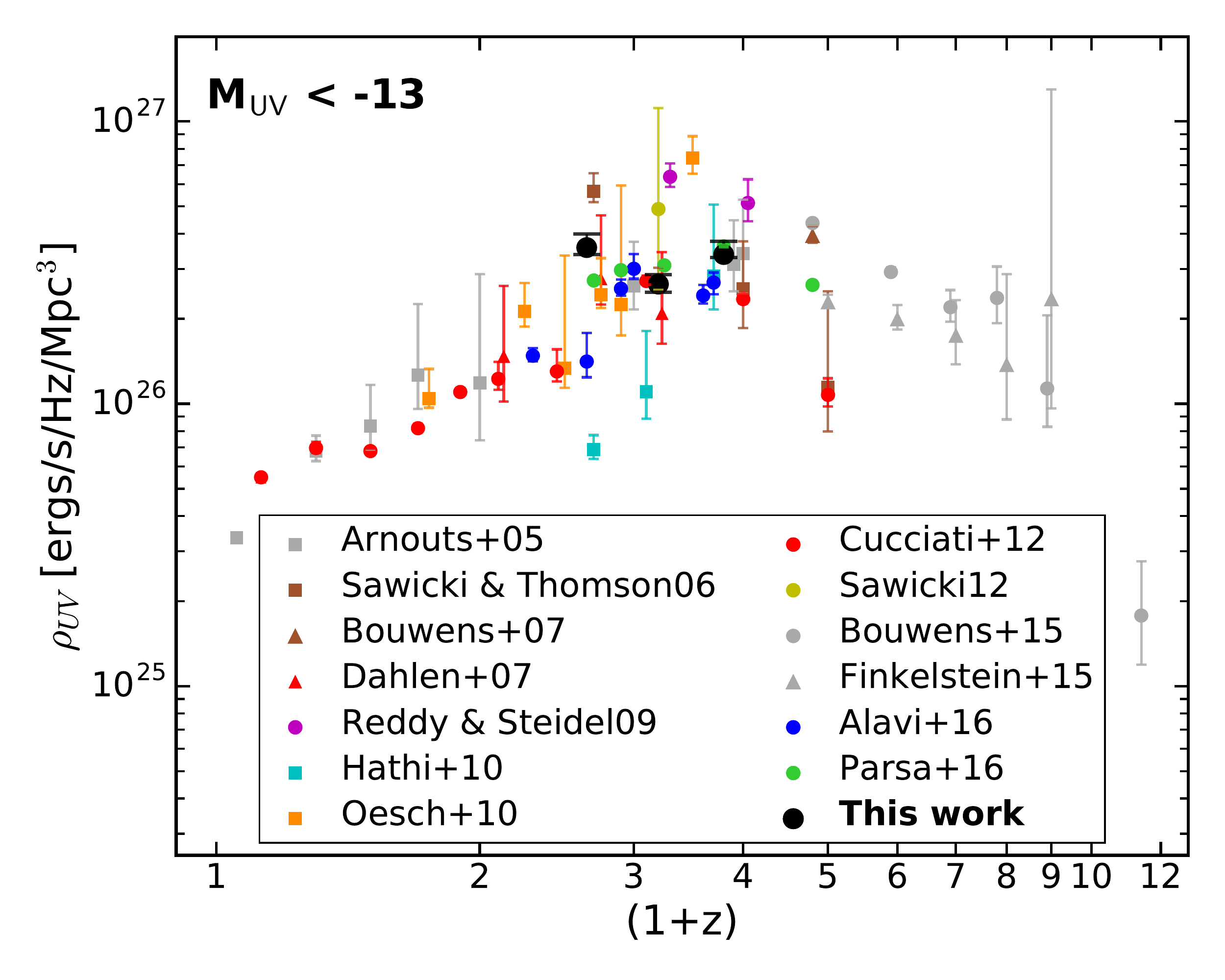}
\caption{Redshift evolution of the observed UV luminosity density (not corrected for dust). Our points are shown in \textit{black} in comparison to various rest-frame UV LFs available in the literature at different redshifts. All the points are derived by integrating the rest-frame UV LFs down to $M_{UV}=-13$ and the errors on the points are estimated using the uncertainties in the LF parameters as reported by the individual references.}
\label{fig:uvld}
\end{figure}

\noindent Table~\ref{tab:uvld} reports the UV luminosity density computed integrating down to a variety of luminosity limits. For this calculation, we use the LF fits derived using the photometric redshift samples, due to their smaller statistical uncertainties as well as coverage down to fainter luminosities. The evolution of the UV luminosity density (not corrected for dust) over redshift is shown in Figure~\ref{fig:uvld}. All points shown were integrated down to $M_{UV}=-13$ according to Equation~\ref{eqn:uvld} in a consistent fashion, using the LF parameters from the cited references along with the reported uncertainties. From $z=0$ to $z=2$, the observed UV luminosity density rises \citep{arnouts05} and peaks around $z\sim2-3$, after which it slightly declines again \citep{finkelstein15, bouwens15}.

Overall, where multiple estimates are available, there is a large scatter in the UV luminosity density. Particularly in the $z\sim1.5$ to $3.5$ range, this scatter is many times larger than the formal errors quoted by some of the surveys, indicating that systematic errors (possibly resulting from the different selection functions and cosmic variance) are not accounted for. Our estimates at $z\sim2-3$ are within 20\% of the two surveys most similar to ours, \citet{alavi16} (using lensed galaxies in HFF and Abell 1689) and \citet{parsa16} (using HUDF without the additional NUV coverage). At the lower redshift $z\sim1.7$, our UV luminosity density estimate is a factor 2.5 and 1.3 higher than \citet{alavi16} and \citet{parsa16}, respectively. This discrepancy is caused due to the high number density we find at the bright end compared to the other two LFs. However, we would like to reemphasize that the coverage area of UVUDF is very small and hence, our result is affected by high cosmic variance.

\begin{deluxetable}{lccc}
\tabletypesize{\footnotesize}
\tablecaption{UV Luminosity Density (not corrected for dust)}
\tablehead{ 
\multirow{2}{*}{Redshift} & \multicolumn{3}{c}{UV Luminosity Density\tablenotemark{a}} \\
\quad & $M<-0.03M_{UV}^\star$\tablenotemark{b} & $M<-13$ & $M<-10$}
\startdata
$z\sim1.65$ & $3.37^{+0.42}_{-0.20}$ & $3.57^{+0.45}_{-0.20}$ & $3.60^{+0.48}_{-0.17}$ \\
$z\sim2.2$ & $2.42^{+0.20}_{-0.18}$ & $2.65^{+0.29}_{-0.15}$ & $2.69^{+0.30}_{-0.15}$ \\
$z\sim3.0$ & $2.99^{+0.33}_{-0.12}$ & $3.38^{+0.35}_{-0.09}$ & $3.43^{+0.40}_{-0.09}$ \\
\vspace{-0.12in}
\tablenotetext{a}{in units of $\times 10^{26}$ ergs/s/Hz/Mpc$^3$}
\tablenotetext{b}{The $M_{UV}^\star$ value used is from our LF fits using the photometric redshift sample, as reported in Table~\ref{tab:pars}.}
\enddata
\label{tab:uvld}
\end{deluxetable}

Using the UV luminosity density, we now compute the cosmic star formation rate density (SFRD) of the Universe. Two main assumptions enter this conversion: $i$) the correction applied to the UV luminosity to account for interstellar dust attenuation, and $ii$) the conversion between UV luminosity and SFR (which depends on, e.g., stellar age, star formation history, and initial mass function, IMF).

We implement the widely used $IRX-\beta$ relation \citep[][hereafter M99]{meurer99} to derive the average UV extinction as a function of the observed UV luminosity. The average $\beta$ for our sample is derived as a function of UV luminosity using the $\beta-M_{UV}$ relation for the appropriate redshift from \citet{kurczynski14}. The resulting $IRX - \beta - M_{UV}$ relation quantifies the dust extinction at the observed UV luminosity. For full details on the applied UV dust correction see Appendix~\ref{appndx:uvdust}. The dust corrected UV luminosity is converted into a star formation rate using the transformations tabulated in \citet{kennicutt12} (which quotes \citealt{murphy11}). The computed SFRDs are reported in Table~\ref{tab:sfrd} for the same luminosity ranges used in Table~\ref{tab:uvld}.

At $z\sim2$, we compare our result to the total intrinsic SFRD computed from the $UV$ and $IR$ data in \citet{madau14}, $\psi_{UV+IR}(z=2.2) = 0.127$ M$_\odot$ yr$^{-1}$ Mpc$^{-3}$. We measure a dust-corrected UV SFRD of $\psi_{UV} = 0.103$ M$_\odot$ yr$^{-1}$ Mpc$^{-3}$ (here, we use the \citealt{kennicutt98} transformation in order to match the \citealt{madau14} analysis) for $M_{UV}< 0.03M_{UV}^\star$, where $M_{UV}^\star$ is the measurement from our rest-frame UV LF fit using the photometric redshift sample. We find that the derived $\psi_{UV}$  is approximately a factor of $\sim 1.2$ lower than the total intrinsic SFRD computed from the $UV$ and $IR$ data in \citet{madau14}. At face value, this result suggests that the correction for dust extinction that we apply is underestimated. 

\begin{deluxetable}{cccc}
\tabletypesize{\footnotesize}
\tablecaption{Star Formation Rate Density (dust corrected)}
\tablehead{ 
\multirow{2}{*}{Redshift} & \multicolumn{3}{c}{SFR Density\tablenotemark{a,b}} \\
\quad & $M<0.03M^\star$\tablenotemark{c} & $M<-13$ & $M<-10$}
\startdata

\cutinhead{Using \citet{meurer99} relation}

$z\sim1.65$ & $0.094^{+0.017}_{-0.008}$ & $0.097^{+0.017}_{-0.008}$ & $0.098^{+0.017}_{-0.008}$ \\
$z\sim2.2$ & $0.066^{+0.008}_{-0.006}$ & $0.070^{+0.010}_{-0.005}$ & $0.070^{+0.010}_{-0.005}$ \\
$z\sim3$ & $0.086^{+0.013}_{-0.005}$ & $0.092^{+0.013}_{-0.003}$ & $0.093^{+0.012}_{-0.004}$ \\

\cutinhead{Using \citet{castellano14} relation}

$z\sim1.65$ & $0.212^{+0.036}_{-0.020}$ & $0.219^{+0.038}_{-0.020}$ & $0.220^{+0.036}_{-0.017}$ \\
$z\sim2.2$ & $0.148^{+0.020}_{-0.013}$ & $0.157^{+0.020}_{-0.014}$ & $0.158^{+0.022}_{-0.011}$ \\
$z\sim3$ & $0.194^{+0.026}_{-0.010}$ & $0.208^{+0.030}_{-0.009}$ & $0.210^{+0.030}_{-0.010}$ \\

\cutinhead{Using \citet{reddy15} relation}

$z\sim1.65$ & $0.120^{+0.022}_{-0.010}$ & $0.125^{+0.022}_{-0.009}$ & $0.125^{+0.023}_{-0.010}$ \\
$z\sim2.2$ & $0.084^{+0.011}_{-0.007}$ & $0.089^{+0.011}_{-0.007}$ & $0.090^{+0.011}_{-0.007}$ \\
$z\sim3$ & $0.110^{+0.016}_{-0.004}$ & $0.118^{+0.017}_{-0.005}$ & $0.119^{+0.018}_{-0.005}$ \\

\vspace{-0.12in}
\tablenotetext{a}{in units of M$_\odot$/yr/Mpc$^3$}
\tablenotetext{b}{using the \citet{kennicutt12} transformations}
\tablenotetext{c}{The $M^\star$ value used is from our LF fits using the photometric redshift sample, as reported in Table~\ref{tab:pars}.}
\enddata
\label{tab:sfrd}
\end{deluxetable}

We can independently check this result using the \ha\ LF. For this calculation, we use the \citet{sobral13} $z\sim2.23$ \ha\ LF, removing the AGN contribution using the $L_{H\alpha}/L^\star_{H\alpha}$ vs. AGN fraction relation presented in  \citet{sobral16}. To account for dust extinction, we use the luminosity dependent dust correction from \citet{hopkins01}, updated for $z\sim 2$ according to the \citet{dominguez13} results. For full details on the applied \ha\ dust correction, see Appendix~\ref{appndx:hadust}.

For a direct comparison, we compute the SFRD by integrating the \ha\ LF down to an \ha\ luminosity corresponding to a SFR of $\sim0.5$ M$_{\odot}$ yr$^{-1}$ (i.e., the SFR corresponding to $0.03M_{UV,z=2}^\star$) and converting the \ha\ luminosity density into SFRD (using \cite{kennicutt98} again, to match the \citealt{madau14} analysis). The resulting SFRD is $\psi_{H\alpha} = 0.116$  M$_\odot$ yr$^{-1}$ Mpc$^{-3}$, more in agreement with the \citet{madau14} $UV+IR$ prediction, and thus, pointing to the dust correction as the main reason for the discrepancy between the SFRD computed from the UV LF alone, and that computed from the $UV+IR$.

\ha\ and UV as star formation indicators have been compared using multi-wavelength studies, both locally \citep[e.g.,][]{lee09,lee11,sanchez12,weisz12,koyama15} as well as at high redshifts \citep[e.g.,][]{wuyts11,shivaei16}. \ha\ has been shown to be a non-biased SFR indicator that agrees well with the total star formation in normal star-forming galaxies. Whereas, UV as a star formation indicator by itself is affected with the problem of dust correction. While it is possible that cosmic variance can impact our UV LF, it is important to note that at the redshift we are considering the bright end of our LF is in very good agreement with other surveys covering large areas. The result, therefore, is not expected to change drastically due to cosmic variance.

\section{Discussion}
\label{sec:discuss}
We further investigate the discrepancies in the volume averaged SFR at $z\sim2$ using the rest-frame UV LF derived in this work and comparing it to the \ha\ LFs available in the literature. We start by computing the star formation rate functions (SFRFs) from both the UV and \ha\ LFs. Similar to the LFs, a SFRF measures the number density of galaxies, but as a function of the star-formation rate, instead of luminosity. Converting the LFs to SFRFs requires transforming the luminosities into a star-formation rate. As before, we correct the UV LF according to typical dust prescription, the M99 $IRX - \beta$ relation (expanded to a $IRX - \beta - M_{UV}$ relation), and the \ha\ LF according the \citet{hopkins01} relation adjusted using \citet{dominguez13} results. Both UV and \ha\ dust-corrected LFs are then converted into SFRFs, using the transformation tabulated in \citet{kennicutt12} (which quotes \citealt{murphy11}). \footnote{Specifically, we use:
\begin{align*}
\mathrm{SFR_{UV} \ [M_\odot / yr]} &=& 10^{-43.35} & \cdot \mathrm{\nu L_{\nu,UV,corr}} \ \mathrm{[erg/s]} \\
    &=& 0.893 \times 10^{-28} & \cdot \mathrm{L_{\nu,1500,corr}} \ \mathrm{[erg/s/Hz]} \\
\mathrm{SFR_{Ha} \ [M_\odot / yr]} &=& 10^{-41.27} & \cdot \mathrm{L_{H\alpha,corr}} \ \mathrm{ [erg/s]} \\
    &=& 5.37 \times 10^{-42} & \cdot \mathrm{L_{H\alpha,corr}} \ \mathrm{[erg/s]}
\end{align*}
}.

\begin{figure}
\centering
\includegraphics[width=0.45\textwidth]{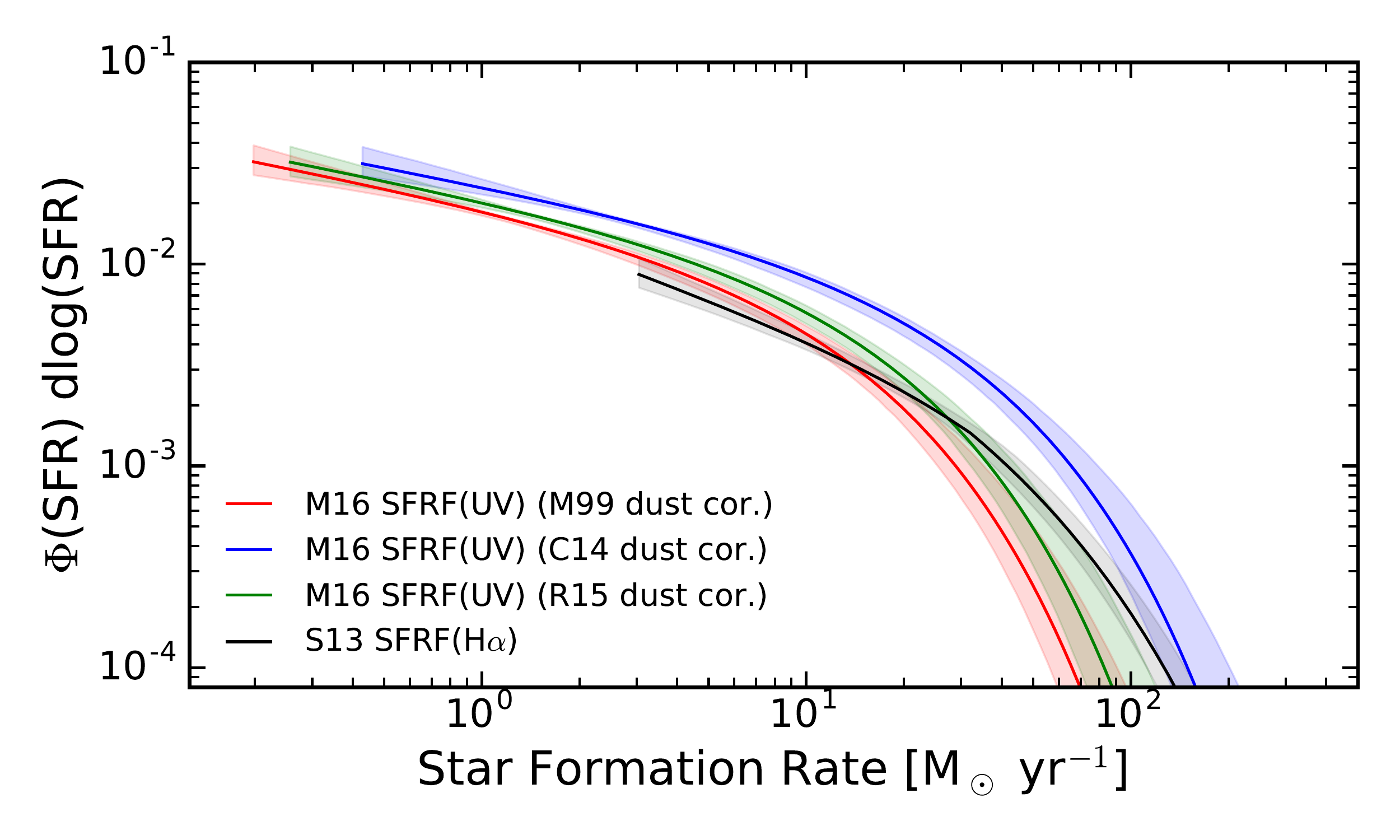}
\caption{Star formation rate functions derived from the UV luminosity function (this work) with dust corrections applied using the generic \citet{meurer99} (in \textit{red}) $IRX-\beta$ relation as well as \citet{castellano14} (in \textit{blue}) and \citet{reddy15} (in \textit{green}) relations, which are calibrated at high redshifts ($z\sim2-3$). The \ha\ star formation rate function derived using the \citet{sobral13} \ha\ LF and corrected using \citet{hopkins01} (updated for high-redshift using \citealt{dominguez13} observations) is plotted in \textit{black}. The UV SFRF corrected using the M99 dust law shows large discrepancy with the \ha\ SFRF, particularly at the bright end. Note that the LFs have only been plotted down to the survey limits. See text for full details on the applied dust corrections.}
\label{fig:SFRF}
\end{figure}

Figure~\ref{fig:SFRF} shows the resulting $z\sim2$ SFRFs, with the SFRF calculated using M99 shown in \textit{red} and the SFRF from \ha\ shown in \textit{black}. The comparison between these two estimates shows that most of the discrepancy originates at the bright end. For SFR $>$30 M$_\odot$ yr$^{-1}$, the \ha\ SFRF estimates for a factor of $\sim$2.5 more sources compared to the UV SFRF. There is clear tension between the rest-frame UV and \ha\ LFs at $z\sim2$, under typical assumptions. Hence, one or more of the assumptions made require additional scrutiny. Recalling the main assumptions that enter this analysis:

\begin{itemize}
\item \textit{Dust}: The observed light (both UV and \ha) needs to be corrected for interstellar dust absorption before the light can be converted to a star formation rate.
\item \textit{Stellar Population Properties}: The intrinsic amount of light emitted from a galaxy (star forming or not) depends on stellar population, age, metallicity, and IMF. Consequently, when interpreting the galaxy light as a star formation rate, one has to assume a stellar population model. This assumption can be broken into finer details: star formation history, stellar age, metallicity, and IMF.
\end{itemize}

It is a not straightforward to investigate all of these assumptions simultaneously with only one measurement each of two observables, rest-frame UV and \ha\ LFs, due to the various degeneracies involved. Here, we individually examine the effects of the most important assumptions: dust and star formation histories.

\subsection{Dust Correction}
Correcting for dust is a key step in going from the observed luminosity to the star formation rate. The $IRX-\beta$ relation in M99 was calibrated using local star-forming galaxies. However, the stellar population properties of high-redshift galaxies may be different from local objects, e.g., they are expected to have lower metallicities and younger ages than their local counterparts. This may cause high-redshift galaxies to have intrinsically bluer UV slopes, and using the M99 relation could underestimate the dust content \citep{wilkins12}. Recently, amendments to the original M99 $IRX-\beta$ relation have been suggested for high redshift galaxies \citep[e.g.,][]{heinis13,castellano14,reddy15}. Here, we test two of the recently suggested prescriptions for $z\sim2$ galaxies: \citet{castellano14} and \citet{reddy15}.

Using a sample of well studied $z\sim 3$ Lyman break galaxies (LBGs), \citet[][hereafter, C14]{castellano14} recently pointed out a systematic offset between the SFR(UV) computed using M99 and those computed from SED fitting. They provide a modification to the $IRX-\beta$ relation by only correcting for the systematic offset: $A_{UV} = 5.32^{+0.41}_{-0.37} + 1.99\beta$, implying a larger correction for dust than using M99. \citet[][hereafter, R15]{reddy15} study a sample of $z\sim 2 $ star forming galaxies with deep optical spectroscopy and multi-wavelength photometry. They fit an $IRX - \beta$ slope of $A_{UV} = 4.48 + 1.84\beta$ for their $z\sim2$ sample, which also implies a slightly higher correction for dust than M99. The difference between these two and the M99 relations is highlighted in Appendix~\ref{appndx:uvdust}.

In Figure~\ref{fig:SFRF}, we  show the SFRFs computed from the UV LF, but assuming the C14 (\textit{blue} curve) and the R15 (\textit{green} curve) $IRX-\beta$ relation. Both of these dust prescriptions reduce the tension at the highest SFR. The C14 relation, in fact, over-corrects the UV SFRF and only agrees with the \ha\ SFRF at the high SFR end (SFR $\gtrsim 80$ M$_\odot$ yr$^{-1}$). This is not entirely surprising when considering that C14 only applied an overall offset to the $IRX - \beta$ relation, which they compute using a sample of high SFR galaxies (SFR $\sim 100$ M$_\odot$ yr$^{-1}$). Ideally, the C14 calibration is only valid for the bright end. The UV SFRF corrected using R15 relation also results in more high-SFR galaxies compared to M99; however, it is still unable to reproduce all of the high SFR galaxies that are recovered in the \ha\ SFRF. Thus, a simple tweaking of the dust prescription is not sufficient to solve the discrepancy, as this prescription only measures the average behavior of a galaxy population. 

The high-SFR end of the SFRF can be also altered artificially because of  the intrinsic scatter in the $IRX - \beta - M_{UV}$ relation. This effect is partially accounted for in the dust extinction as calculated in Appendix~\ref{appndx:uvdust}, where the scatter between $\beta$ and $M_{UV}$ is considered (by the $0.2 \ln{10} b^2 \sigma_\beta^2$ term in the $A_{UV} - \beta$ conversion). This correction assumes that the scatter is constant with luminosity and symmetric with respect to the best-fit relation.  However, studies have shown that in fact this is not the case, and the scatter increases towards fainter magnitudes, and the distribution around the best-fit becomes skewed, because of the dust--free limit on $\beta$  \citep[e.g.,][]{bouwens12, alavi14, kurczynski14}.  Because of this effect, some faint galaxies may in reality have large dust corrections. A few of these objects would be sufficient to affect the bright end of the SFRF, given its exponential fall-off. A proper treatment of the scatter could then reduce the tension between the UV and \ha\ SFRFs.

There is considerable evidence for the existence of a population of the so-called ``Dusty SFGs" (DSFGs; dusty star-forming galaxies), both locally \citep{goldader02, burgarella05, buat05, howell10, takeuchi10} and at high redshifts \citep{reddy10, heinis13, casey14, ivison16}. The DSFGs are a population of galaxies that have high star-formation rates (SFR $>50$ M$_\odot$ yr$^{-1}$) and high IR luminosities ($L_{IR} \gtrsim$ few $\times 10^{11} L_\odot$). However, due to their high dust content, they have a high $IRX$ ($L_{IR}/L_{UV}$) that is offset from the nominal $IRX - \beta$ relation. These galaxies are faint in the UV not because they are intrinsically faint and have low dust content, but instead because they are intrinsically bright and are highly obscured by dust. A generic $IRX-\beta$ relation would underestimate the dust content for these galaxies and hence, would result in a deficit of high SFR sources (SFR $>50$ M$_\odot$ yr$^{-1}$). The shortage of sources in the UV SFRF at the high SFR end can plausibly be explained by these objects.

\subsection{Star Formation Histories}
The conversion between light (either \ha\ or UV) and SFR is another key step in the calculation of the SFRFs, which depends critically on the age of the star-burst, and therefore on the specific star formation history. Until this point, we have assumed the conversion from \citet{kennicutt12}, which implicitly assumes that the SFR has been constant for at least 100 Myr.  If this is not the case, however, the \citet{kennicutt12} conversion is not justified, and we have to take into account the fact that the rest-frame UV and the \ha\ luminosities are sensitive to star formation occurring over different timescales.

\begin{figure}
\centering
\includegraphics[width=0.45\textwidth]{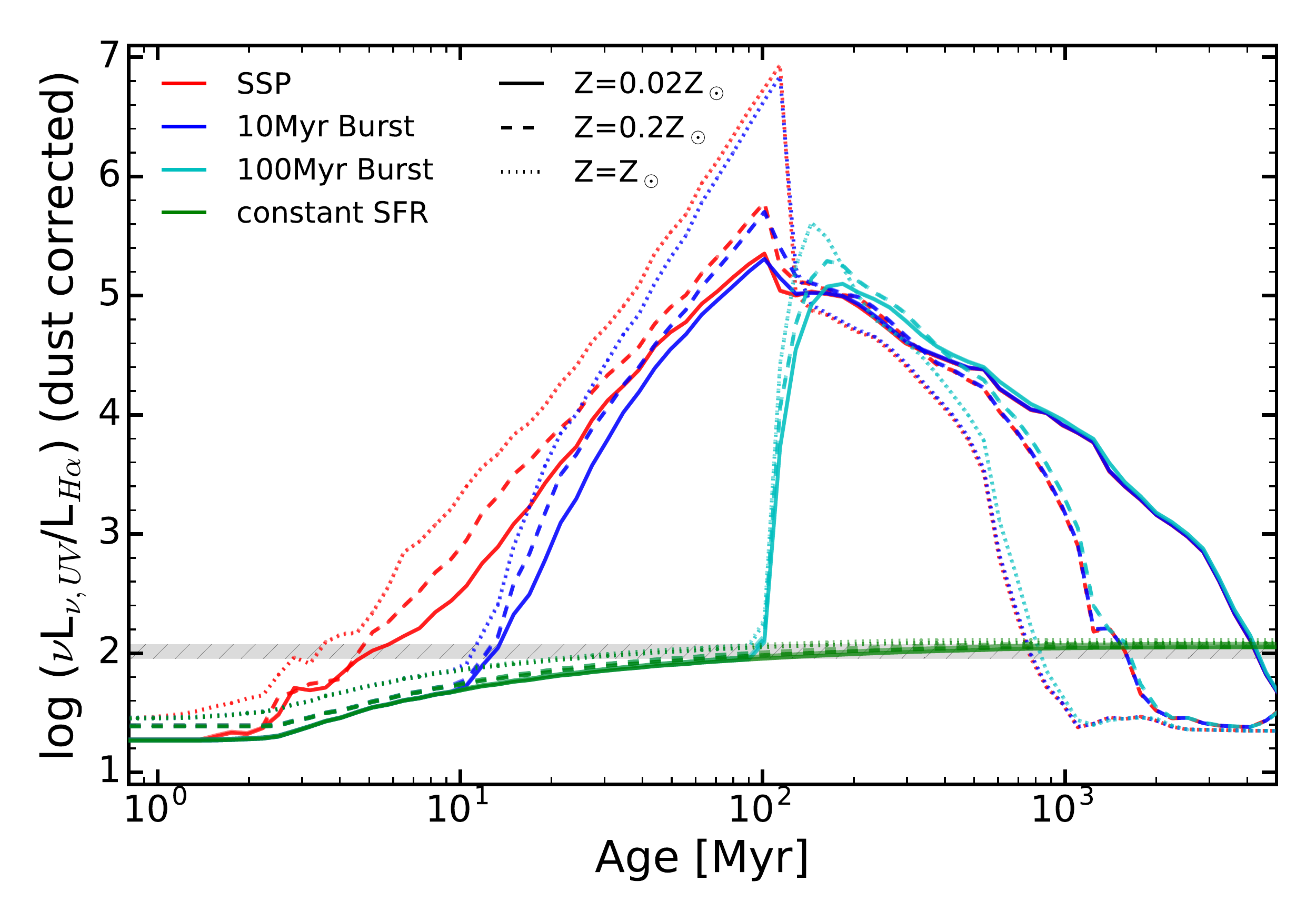}
\caption{The UV-to-\ha\ ratio plotted as a function of age since the onset of star formation for a range of star formation histories (single stellar population, single bursts, rising, and constant star formation) computed using \citet{bc03} models. The shaded band shows the expected log$_{10}[\nu L_\nu (1500)/L_{H\alpha}]$ value for constant star formation, after accounting for a range of metallicities as well as different IMFs. See text for description of the models.}
\label{fig:uv_vs_ha_ratio}
\end{figure}

In particular, the rest-frame UV is sensitive to star formation occurring over $\sim$100s of Myrs, whereas the \ha\ is sensitive to star formation over $\sim$ 10s of Myrs. The brightness of a galaxy in the two indicators (i.e., \ha\ and FUV) depends on the recent star formation history. Figure~\ref{fig:uv_vs_ha_ratio} shows the evolution of the UV-to-\ha\ ratio for a variety of star formation histories. The \ha\ output from a galaxy drops on much shorter timescales, after the end of a burst, compared to the non-ionizing UV, which takes longer to react, because of the longer lived B and A stars that still produce UV photons, but have very little ionizing output. For a constant star formation history, the rates at which massive stars are formed and die reach an equilibrium after approximately 100 Myr, and therefore the UV-to-\ha\ ratio tends to a constant value (log$_{10}[\nu L_\nu (1500)/L_{H\alpha}] \sim 2$); whereas, a burst of  star formation would cause the UV-to-\ha\ ratio to scatter towards higher values.

Figure~\ref{fig:uv_vs_ha_ratio} shows how the UV-to-\ha\ ratio evolves for \cite{bc03} models with \cite{salpeter55}, \cite{kroupa01} as well as \cite{chabrier03} IMFs and metallicities of $Z/Z_\odot=0.02, 0.2, 1$ with three different star formation histories: single instantaneous burst (SSP), short bursts of star formation (10Myr and 100Myr), and constant star formation. The main impact of the different IMFs on the UV-to-\ha\ ratio is limited to the values of their slopes at the high-mass end, since both \ha\ and UV are sensitive to hot, massive stars. Moreover, this difference in IMF slopes at the high masses is small enough between \citet[][$\alpha=-2.35$]{salpeter55}, \citet[][$\alpha=-2.3$]{kroupa01}, and \citet[][$\alpha=-2.3$]{chabrier03} IMFs that the resulting variation in the UV-to-\ha\ ratio is within the linewidths of the curves in Figure~\ref{fig:uv_vs_ha_ratio}. For models with constant star formation histories, we compute the range of expected UV-to-\ha\ ratios at an age of 100Myr (the generic assumption in SFR conversions, e.g., \cite{kennicutt12}). The horizontal band in Figure~\ref{fig:uv_vs_ha_ratio} shows this expectation from the constant star formation history case.

The effect of the star formation histories (SFHs) on the UV-to-\ha\ ratio has been studied both in simulations \citep{shen13, hopkins14, dominguez15} as well as observations \citep{boselli09, finkelstein11, lee11, lee12, weisz12, guo16}. \citet{dominguez15} used SFHs derived from hydro-dynamical simulations to study the variation of  the  UV-to-\ha\ ratio, which they suggest is a useful observable to quantify the ``burstiness'' of a galaxy's SFH. Using their simulated galaxies they find that galaxies with low stellar masses ($M_\star \lesssim 10^9 M_\odot$) are dominated by bursty SF, and have a higher mean value and scatter of the  UV-to-\ha\ ratio compared to more massive galaxies. This is a result of energy feedback from star formation being more efficient in low--mass galaxies \citep{somerville15}. 

We can study the trend between the volume-averaged UV-to-\ha\ ratio and the stellar mass by comparing the dust-corrected UV and \ha\ LFs. To do so, we need to be able to associate a given luminosity (either \ha\ or UV) to its corresponding halo mass. We use a standard  ``abundance matching"  technique \citep[e.g.,][]{conroy06,guo10,moster10,trenti10,tacchella13,mason15} to associate galaxies with given number densities to their corresponding  dark matter halos. The implicit assumption in this step is that there is only one galaxy per dark matter halo.  In practice, we find a relation between the halo mass ($M_h$) and the observed luminosity (e.g., $L_{UV}$) by solving the following equations: 

\begin{equation}
\begin{split}
\int_{M_h}^\infty n(M'_h,z=2.2) dM'_h =& \int_{L_{UV}}^\infty \phi(L'_{UV}) dL'_{UV} \\ 
=& \int_{L_{H\alpha}}^\infty \phi(L'_{H\alpha}) dL'_{H\alpha}
\label{eqAM}
\end{split}
\end{equation}

\noindent where $n(M_h,z)$ is the analytical dark--matter halo mass function from \citep{sheth01} computed at $z=2$, $\phi(L_{UV})$ is the $z\sim2$ UV LF from this work, and $\phi(L_{H\alpha})$ is the $z=2.23$ \ha\ LF from \citet{sobral13}.  By solving Equation~\ref{eqAM}  we derive the  UV and \ha\ luminosities that correspond to a given dark matter halo mass, and thus the UV-to-\ha\ ratio that corresponds to that halo mass. We limit this analysis only to luminosities down to which our $z\sim2$ UV LF sample extends. This does involve extrapolating the \ha\ LF $1 \ dex$ below their observation limit. Also, note that in order to compare to intrinsic flux ratio, the observed luminosities still need to be corrected for dust. We explore two dust relations (M99 and R15) from the previous section.

\begin{figure}
\centering
\includegraphics[width=0.45\textwidth]{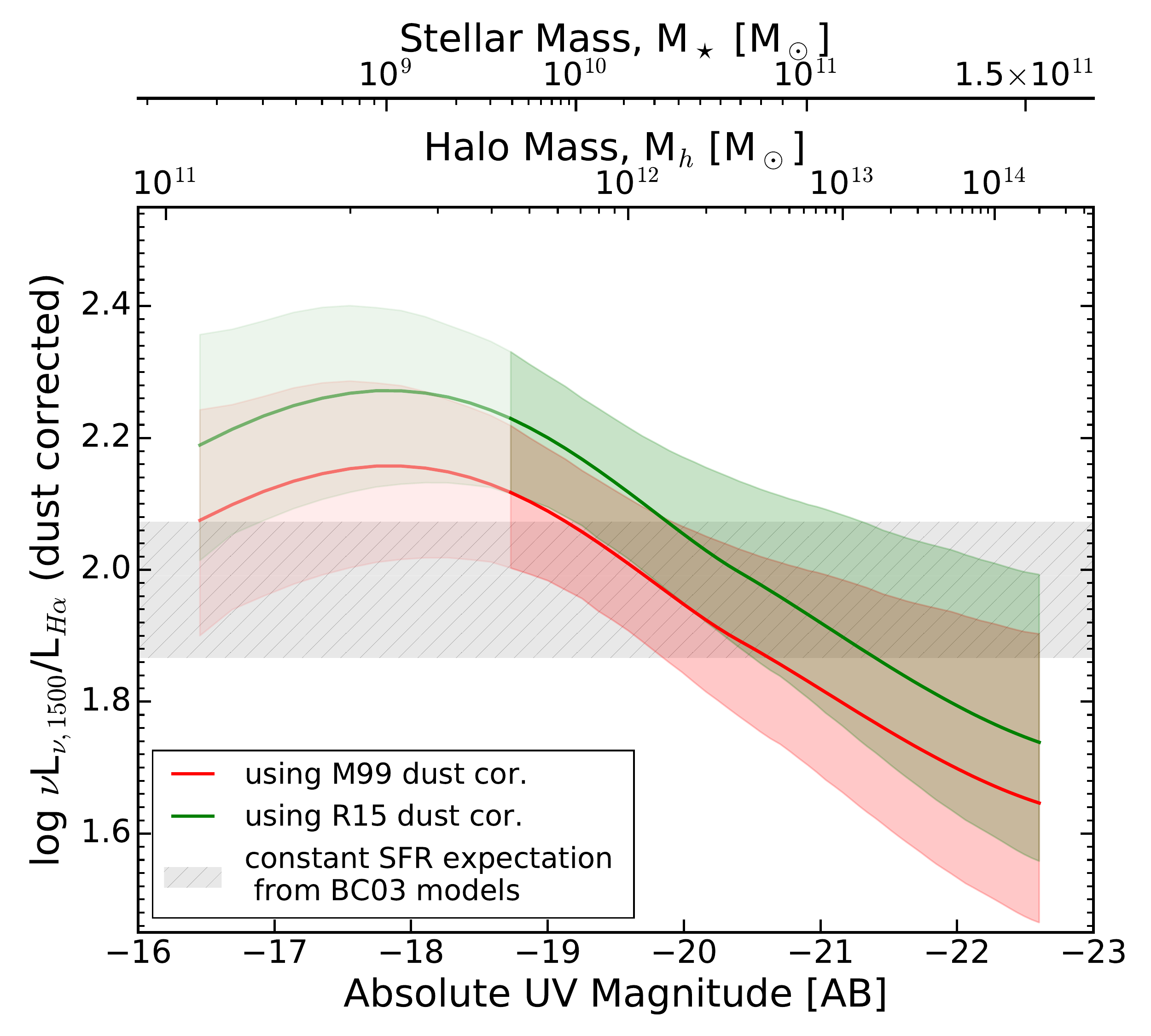}
\caption{The observed volume-averaged UV-to-\ha\ ratio plotted as a function of the observed UV luminosity using the empirical relations derived from abundance matching the $z\sim2$ UV LF and \citet{sobral13} \ha\ LF, using two UV dust prescriptions: \citet{meurer99} (\textit{red} curve) and \citet{reddy15} (\textit{green} curve). The horizontal band again shows the expected log$_{10}[\nu L_\nu (1500)/L_{H\alpha}]$ value for constant star formation, including a variety of IMFs and metallicities (see text for details). The \textit{abundance-matched} DM halo mass is plotted on the top axis. Corresponding stellar masses are computed using the $z=2.2$ stellar mass-halo mass relation from \citet{behroozi13} and plotted on a parallel axis. The shaded regions show the $1\sigma$ confidence regions for the measured empirical relation. The darker shaded bands show the range where both UV and \ha\ LFs have observations; whereas, the lighter shaded bands show the range when UV LFs have observations, but the \ha\ LF has been extrapolated.}
\label{fig:uv_vs_ha_mass}
\end{figure}

Figure~\ref{fig:uv_vs_ha_mass} shows the main result of this analysis. The horizontal hatched region from Figure~\ref{fig:uv_vs_ha_ratio} is shown again to highlight the expected range of values for log$_{10}[\nu L_\nu (1500)/L_{H\alpha}]$ assuming a constant SFH and a range of metallicities ($Z/Z_\odot=0.02 - 1$) and different IMFs \citep{salpeter55,kroupa01,chabrier03}. The derived volume-averaged UV-to-\ha\ ratio is not observed to be constant as a function of halo mass, but rather it increases above the value expected for constant SFH, as one moves towards the lower halo-mass end ($M_h \lesssim10^{12}M_\odot$). As seen in Figure~\ref{fig:uv_vs_ha_ratio}, the impact of variation in IMF and/or metallicities on the UV-to-\ha\ ratio is small compared to the variations in the SFH. At face value, the increased ratio at low masses indicates a larger contribution by starbursting objects to the average population of galaxies, as suggested by the larger scatter predicted by \citet{dominguez15}.  The halo mass where this effect seems to be important is  below $\sim 5\times10^{11}$ M$_{\odot}$, which would correspond to a stellar mass of $\sim 5 \times 10^{9}$ M$_{\odot}$, assuming the stellar mass to halo mass relation from \citet{behroozi13}. We also provide polynomial fits to the empirical UV-to-\ha\ relation in Table~\ref{tab:emp_fits}.

It is important to note that the measured empirical UV-to-\ha\ ratio from abundance matching in Figure~\ref{fig:uv_vs_ha_mass} is very sensitive to the applied dust relation. However, they all exhibit the elevated UV-to-\ha\ ratio. The abundance matching technique matches the cumulative UV and \ha\ LFs and hence, it is also sensitive to the systematic differences at the bright end. As noted in the previous section, there is a distinct possibility of the presence of DSFGs in the UVUDF sample that are not being corrected for dust properly by the applied $IRX-\beta$ relation. This impacts the bright end more significantly than the faint-end, due to low number statistics. Changing the bright end of the dust-corrected UV LF would offset the UV-to-\ha\ curves in Figure~\ref{fig:uv_vs_ha_mass} vertically by a significant amount, while changing the overall shape of the curves only  minimally.  Thus the upturn in the UV-to-\ha\ ratio at $M_h \lesssim10^{12}M_\odot$ is preserved, although the characteristic stellar mass where this becomes important is somewhat dependent on the specific dust correction used in the analysis. 

Figure~\ref{fig:uv_vs_ha_mass} clearly shows that the observed trend in the UV-to-\ha\ ratio is not constant with the observed UV luminosity and is inconsistent with  constant star formation rate at all luminosities, even after accounting for a range of metallicities as well as different canonical IMFs. We also considered the possibility of a non-universal IMF, as introduced by \citet[][]{weidner05, pflamm07, pflamm09, weidner11}. Based on statistical arguments, \citet[][]{weidner05} produced an integrated galactic initial mass function (IGIMF) which steepens in galaxies with lower SFRs, which could then reproduce a trend similar to what is observed in  Figure~\ref{fig:uv_vs_ha_mass}. However, the main impact of the SFR-dependent IGIMF occurs at  SFR much lower than those probed by the current analysis \citep[$\lesssim 10^{-2}$ M$_\odot$ yr$^{-1}$;][]{pflamm09}. In the SFR range ($\gtrsim 0.3$ M$_\odot$ yr$^{-1}$) probed here, the  IGIMF is in fact constant with SFR, and thus cannot account for the observed trend in the UV-to-\ha\ empirical relation by itself.

Lastly, the \ha\ extinction correction we use is rather uncertain and hence, we explore whether the observed trend can be explained uniquely with dust. If dust were the only cause, the required \ha\ extinction correction would have to be nearly constant with \ha\ luminosity ($A(H\alpha)\sim0.5$) and thus stellar mass. The latter constraint is, however, inconsistent with observational results.  Using a compilation of results from the recent literature, \citet{price14} show that the dust extinction increases  for brighter, higher-mass galaxies, consistent with earlier results by \citet{garn10}.

\begin{deluxetable*}{ccccccccc}
\tabletypesize{\footnotesize}
\tablecaption{Polynomial fits to the empirical UV-to-\ha\ relation\tablenotemark{a}}
\tablehead{UV Dust correction & $a_6$ & $a_5$ & $a_4$ & $a_3$ & $a_2$ & $a_1$ & $a_0$ & $\sigma$\tablenotemark{b}}
\startdata

\cutinhead{$X = (M_\mathrm{UV,AB} - 20)$\tablenotemark{c}}

\cite{meurer99} & 1.46e-04 & -2.13e-05 & -2.49e-03 & -6.37e-03 & 2.21e-03 & 1.41e-01 & 1.95 & 0.154 \\
\cite{reddy15} & 1.46e-04 & -1.19e-05 & -2.51e-03 & -6.60e-03 & 1.92e-03 & 1.46e-01 & 2.06 & 0.153 \\

\cutinhead{$X = \log{(M_h / 10^{12}M_\odot)}$\tablenotemark{d}}

\cite{meurer99} & -7.61e-03 & 5.58e-02 & -1.42e-01 & 1.28e-01 & 5.19e-02 & -2.99e-01 & 2.01 & 0.154  \\
\cite{reddy15} & -7.67e-03 & 5.66e-02 & -1.44e-01 & 1.31e-01 & 5.28e-02 & -3.10e-01 & 2.12 & 0.153 \\

\vspace{-0.12in}
\enddata
\tablenotetext{a}{The empirical relation: log$(\nu L_\nu (1500)/L_{H\alpha}) = \displaystyle\sum_{i=0}^{N} a_i  X^i$}
\tablenotetext{b}{$1\sigma$ scatter about the mean relation}
\tablenotetext{c}{$M_\mathrm{UV,AB}$ range = (-16.45,-22.60)}
\tablenotetext{d}{$\log{(M_h /M_\odot)}$ range = (11.05,14.3)}
\label{tab:emp_fits}
\end{deluxetable*}

\section{Conclusions}
\label{sec:conclusions}
NUV coverage of the \textit{Hubble} UDF provided by the UVUDF enables for LBG dropout and photometric redshift selection of galaxies near the peak of cosmic star formation ($z\sim2-3$). Additionally, it also enables the study of their rest-frame UV properties and consequently, their star formation properties. Here, we present the rest-frame 1500\AA\ UV LFs for F225W ($z\sim1.65$), F275W ($z\sim2.2$), and F336W ($z\sim3$) dropout galaxies in the UVUDF selected by the LBG dropout criteria as well as by their photometric redshifts. We develop and execute a suite of completeness simulations to properly correct the effective volumes when fitting the LFs.

Overall, our best-fit rest-frame UV LFs are in good agreement with the recent results from \citet{parsa16}. We measure faint-end slopes that are within the errors compared to other blank-field surveys \citep[such as,][]{hathi10,oesch10}. There is a striking discrepancy between our results and those from the \citet{alavi16} analysis of lensed galaxies. However, the steep faint-end of the LF measured from lensed galaxies could be the result of systematic uncertainties in the lensing modeling. These systematics are particularly important for the most magnified, i.e., intrinsically faintest,  sources -- those that contribute most to the measurement of $\alpha$ \citep{bouwens16}. 

At $z\sim2.2$, using our F275W dropout sample which covers the absolute UV magnitude range from $-22.00$ to $-17.97$ AB (where the effective volume correction drops to 25\% of that are the bright end), we measure a faint-end slope of $\alpha=-1.31^{+0.32}_{-0.75}$. This is well in agreement with the corresponding photometric redshift sample, which covers a range of $-22.00$ to $-16.30$ AB in absolute UV magnitude, going $\sim1.5$ magnitudes deeper. When compared with results from the literature, we find good agreement with \citet{parsa16} as well as \citet{oesch10} and \citet{sawicki12}, given the uncertainties on their result. At $z\sim1.65$, our best-fit LF estimates a higher number density at the bright end in comparison to other results from the literature. However, due to the small area covered by the UVUDF, this sample is affected by high cosmic variance. For both F275W and F336W dropouts, the LFs measured from the LBG samples agree with those measured from their corresponding photometric redshift samples, within the uncertainties.

We find an observed UV luminosity density (at $M_{UV}<-13$) that is consistent within 20\% of both \citet{alavi16} and \citet{parsa16} at $z\sim2-3$. We apply the \citet{meurer99} dust relation to correct the UV luminosities and compute the star formation rate density (SFRD) and find a factor of 2 discrepancy when compared to the total intrinsic star formation rate from $UV+IR$ observations \citep{madau14}. This discrepancy is absent when using a $z\sim2$ \ha\ LF \citep{sobral13} to compute the SFRD; thus, pointing to the dust correction as the main reason for the discrepancy.

We compute the SFRF from the rest-frame UV LF using the generic M99 dust correction. The SFRF corrected according to the M99 relation failed to recover a factor of $\sim2.5$ high SFR ($> 30$ M$_\odot$ yr$^{-1}$) sources compared to the \ha\ SFRF. We find that using the M99 dust correction, which is calibrated using local galaxies, underestimates the dust content in the high-redshift ($z\sim2$) star-forming galaxies. Using relations calibrated at high-redshift such as \citet{castellano14} and \citet{reddy15}, reduces the tension. However, a straightforward tweaking of the $IRX-\beta$ relation is not sufficient to fully resolve the tension. One possibility is the presence of very dusty SFGs, that would not be properly corrected by the $IRX-\beta$ relation because of their offset from the average relation.

Another factor affecting the differences between the UV and \ha\ LFs is the burstiness of star formation in galaxies. We use \textit{abundance matching} of the rest-frame UV and \ha\ LFs to compute a volume-averaged UV-to-\ha\ ratio -- an indicator of ``burstiness" in galaxies. We find an increasing UV-to-\ha\ ratio towards low halo masses. We conclude that this trend could be due to a larger contribution from starbursting galaxies at lower masses compared to the high-mass end. This trend is consistent with the expectation from hydrodynamical simulations.

\acknowledgments
We thank the anonymous referee for their comments which helped improve the quality of the manuscript. We would like to thank Dan Coe for help with implementing the photometric redshift algorithm (BPZ) into the completeness simulations. Support for HST Program GO-12534 was provided by NASA through grants from the Space Telescope Science Institute, which is operated by the Association of Universities for Research in Astronomy, Inc., under NASA contract NAS5-26555. M.R. also acknowledges support from an appointment to the NASA Postdoctoral Program at Goddard Space Flight Center. M.B.K. acknowledges support from the National Science Foundation (grant AST-1517226) and from NASA through HST theory grants (programs AR-12836, AR-13888, AR-13896, and AR-14282) awarded by the Space Telescope Science Institute (STScI), which is operated by the Association of Universities for Research in Astronomy (AURA), Inc., under NASA contract NAS5-26555.

\appendix

\section{Completeness Functions}
\label{appndx:comp}
\begin{figure}
\centering
\includegraphics[width=0.48\textwidth]{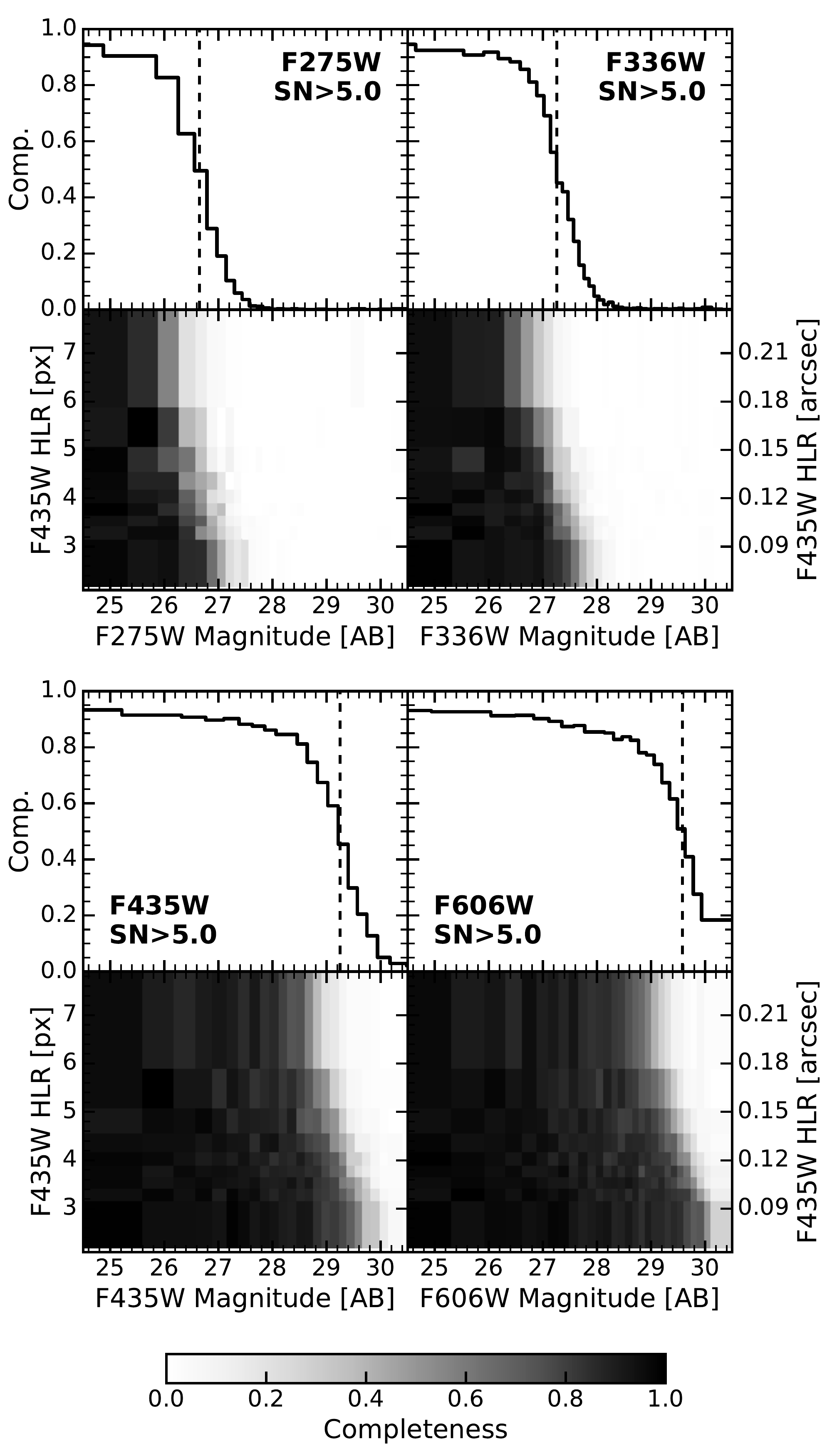}
\caption{Completeness functions for F275W, F336W, and F435W filters (from left to right) for sources detected at $>5\sigma$. The top panel shows the completeness as a function of magnitude only, where as in the bottom panel, completeness is plotted as a function of magnitude and galaxy size. The dashed lines show where the completeness drops to 50\%. The F435W (B-band) half-light radius (HLR) is used as an proxy for galaxy size. From the bottom panels, it is evident that galaxy size has a significant impact on the completeness -- extended sources are missed more often than the more compact ones, even when they have the same magnitude.}
\label{fig:comp}
\end{figure}

From our simulations, the completeness functions are computed using the fraction of recovered sources as a function of observed magnitude as well as galaxy size. We use the  B$_{435}$ half-light radius as a proxy for the galaxy size. These completeness functions are marginalized over all sizes to obtain completeness just as a function of magnitude, which is used to define the survey magnitude limits. These functions are only used to set the survey magnitude limits when computing the colors of sources not detected in the dropout filter. The effective volumes for computing the LF are fully corrected using the selection functions. Figure~\ref{fig:comp} shows the completeness functions for the F275W, F336W, and F435W filters. Note that the compact sources have a higher recovery fraction at fainter magnitudes compared to more extended sources.

\section{UV Dust correction}
\label{appndx:uvdust}
For correcting the observed UV magnitudes, we use the dependence of dust extinction ($A_{UV}$) on the UV slope $\beta$, also known as the $IRX-\beta$ relation. Through the analysis, we implement the $IRX-\beta$ relation ($A_{UV} = a + b \cdot \beta$). We use multiple published fits for this relation: (i) the original \citet[][; M99]{meurer99} calibrated using local star-forming galaxies: $[a,b] = [4.43, 1.99]$, (ii) the \citet{castellano14} calibration using high-SFR LBGs at $z\sim3$: $[a,b] = [5.32^{+0.41}_{-0.37}, 1.99]$, and (iii) the \citet{reddy15} calibration using $z\sim2$ star-forming galaxies: $[a,b] = [4.48,1.84]$.

Furthermore, the UV slope $\beta$ parameters are estimated as a function of the observed absolute UV magnitudes with a $\beta-M_{UV}$ relation. The distribution of $\beta$ for the $z\sim2$ galaxies as a function of the absolute UV magnitudes is assumed using a parametric form, following \citet{trenti15} and \citet{mason15}:

\begin{equation*}
  <\beta> = 
	  \begin{cases} 
      (\beta_{M_0}(z) -c) \exp{\left[-\frac{\frac{d\beta}{dM_{UV}}(z) [M_{UV} - M_0]}{\beta_{M_0}(z)-c} \right]} + c, \\
      \hphantom{-------------} \mathrm{if} M_{UV} \geq M_0 \\
      \frac{d\beta}{dM_{UV}}(z) [M_{UV} - M_0] + \beta_{M_0}(z),\\
      \hphantom{-------------}\mathrm{if} \ M_{UV} < M_0 \\
   \end{cases}
\end{equation*}

\noindent where $c$ is the \textit{dust-free} $\beta$. This relation avoids the unphysical negative values of $A_{UV}$, while also avoiding unphysical discontinuities near magnitudes where the relation approaches \textit{dust-free} $\beta$. The parameters $M_0$, $\beta_{M_0}$, $d\beta / dM_{UV}$ and $\sigma_\beta$ define the $M_{UV} - \beta$ relation and are constrained observationally. We use the results from \citet{kurczynski14}, who derive this relation for $1<z<8$ galaxies using the UVUDF. For $z\sim2$, the applied values are $M_0=-19.5$, $\beta_{M_0}=-1.71$, $d\beta / dM_{UV}=-0.09$, and $\sigma_\beta =0.36$.

Moreover, assuming a Gaussian distribution of $\beta$ with a dispersion $\sigma_\beta$ gives the average extinction: $<A_{UV}> = a + 0.2 \ln{(10)} b^2 \sigma^2_\beta + b <\beta>$, where $b$ is the slope of the $IRX-\beta$ relation \citep{tacchella13,mason15}.

\begin{figure}
\centering
\includegraphics[width=0.45\textwidth]{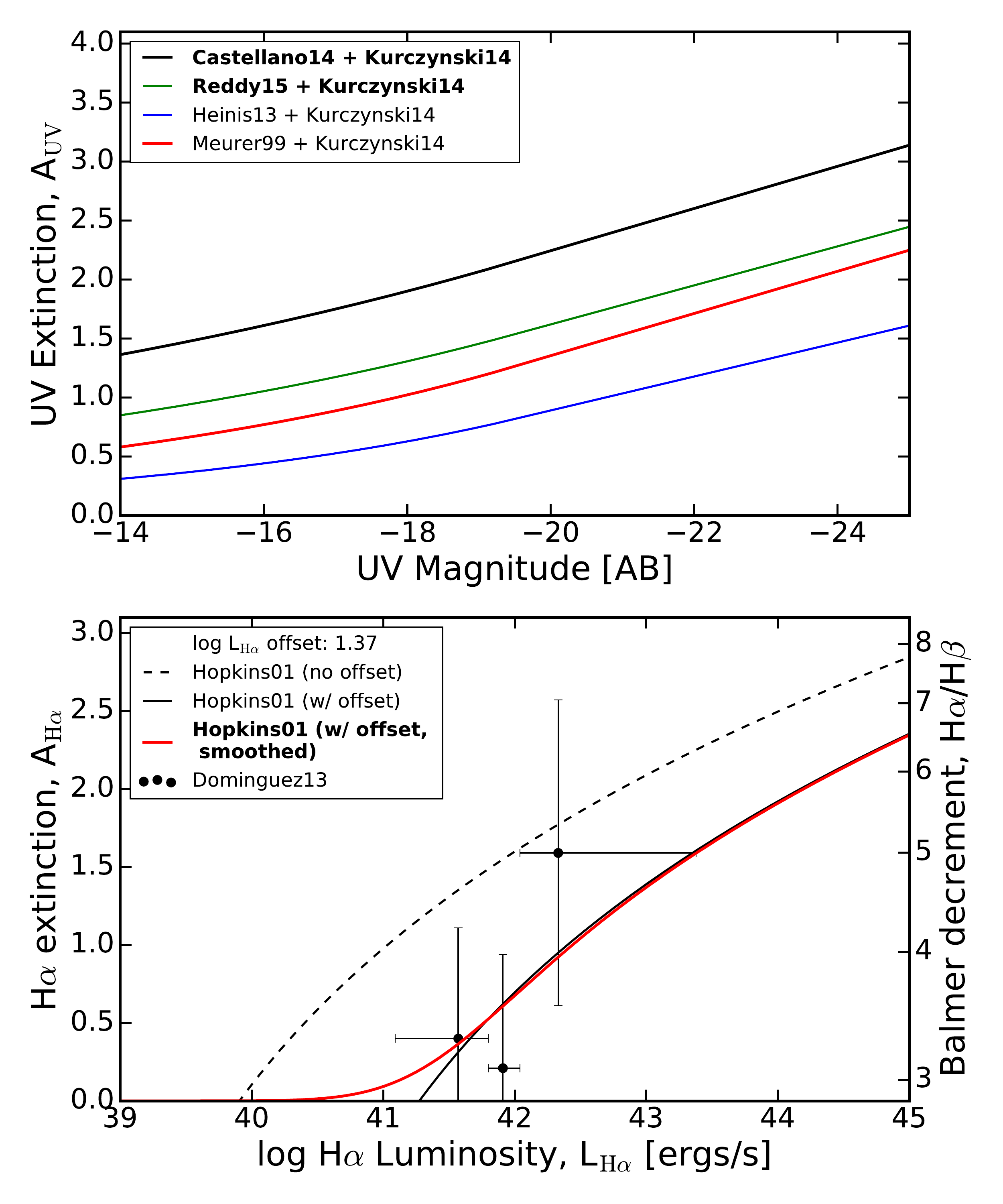}
\caption{Applied dust extinction correction for our analysis. (Top panel) The dust extinction in rest-frame UV as a function of the observed UV absolute magnitude. The widely used \cite[][M99]{meurer99} relation was calibrated using local star forming galaxies, where as the \citet{castellano14} modifies the M99 relation for higher redshift galaxies. (Bottom panel) The dust extinction in the \ha\ line as a function of the observed \ha\ luminosity. The \citet{hopkins01} relation for local galaxies is shown in \textit{dashed black}, whereas the \textit{solid} curves show the same relation updated for $z\sim2$ galaxies according to the \citet{dominguez13} observations. See text for full details.}
\label{fig:dust}
\end{figure}

\section{\ha\ Dust correction}
\label{appndx:hadust}
The observed \ha\ luminosities are corrected for dust extinction by applying the luminosity dependent dust extinction ($A_{H\alpha}$) reported by \citet{hopkins01}. They derive a SFR dependent reddening using a composite of UV, \ha\ emission line and FIR data. Most importantly, their dust correction can be applied as a function of \textit{attenuated} SFR (or light). This is crucial for the analysis here, since the goal is to correct the \ha\ LF for dust as a function of the \textit{observed} \ha\ luminosity.

However, the \citet{hopkins01} relation is derived for local galaxies. In order to scale the relation to match the dust properties of galaxies at $z\sim2$, we apply a shift such that it matches the \citet{dominguez13} observations at $0.7<z<1.5$. This is motivated by the fact that a single relation holds even at higher redshifts, provided that the overall increased star formation (at higher redshift) is accounted for \citep{sobral12}. This can be done by applying an offset in the observed luminosity for the $A_{H\alpha}-L_{H\alpha}$ relation. This can also be interpreted as the typical dust extinction in \ha\ not depending on the absolute star formation rate, but rather a relative dependence -- how bright (or star-forming) a source is relative to the rest of the galaxy.

Following this justification, we apply an offset of $L_{H\alpha} = 10^{1.37}$ to adjust the \citet{hopkins01} local relation to match the Balmer decrement observations of star-forming galaxies at $0.7<z<1.5$ from \citet{dominguez13}. We also apply a smoothing of 0.5 mag to avoid any discontinuities, as these would in turn, create unphysical discontinuities in the SFRFs. The bottom panel of Figure~\ref{fig:dust} shows the dust correction applied to \ha\ luminosities for our analysis.

\begin{deluxetable}{cccc}
\tablecaption{Rest-frame UV LFs measured via the $V_{max}$ estimator\tablenotemark{a}}
\tablewidth{0pt}
\tablehead{
\colhead{Redshift}  & 
\colhead{$M_{1500}$} & 
\colhead{N\tablenotemark{b}} & 
\colhead{$\phi$ [$\times 10^{-3}$ Mpc$^{-3}$ mag$^{-1}$]\tablenotemark{c}}
}
\startdata

\cutinhead{LBG Dropout samples}

\multirow{6}{*}{\parbox{1.8cm}{F275W \\ dropouts \\ $z\sim2.2$}}
 & $-18.22$ & 14 & $ 6.659 _{-1.495} ^{+4.158}$ \\
 & $-18.72$ & 11 & $ 3.447 _{-0.873} ^{+2.428}$ \\
 & $-19.22$ & 15 & $ 3.636 _{-0.778} ^{+2.162}$ \\
 & $-19.72$ & 11 & $ 2.364 _{-0.589} ^{+1.639}$ \\
 & $-20.22$ &  5 & $ 1.049 _{-0.388} ^{+1.079}$ \\
 & $-20.72$ &  2 & $ 0.426 _{-0.249} ^{+0.693}$ \\
 \hline
 
\multirow{9}{*}{\parbox{1.8cm}{F336W \\ dropouts \\ $z\sim3$}}
 & $-17.62$ & 29 & $10.461 _{-1.618} ^{+4.499}$ \\
 & $-18.12$ & 28 & $ 6.492 _{-1.018} ^{+2.832}$ \\
 & $-18.62$ & 36 & $ 6.526 _{-0.904} ^{+2.514}$ \\
 & $-19.12$ & 34 & $ 4.953 _{-0.703} ^{+1.956}$ \\
 & $-19.62$ & 34 & $ 4.216 _{-0.599} ^{+1.666}$ \\
 & $-20.12$ & 16 & $ 1.844 _{-0.381} ^{+1.060}$ \\
 & $-20.62$ & 13 & $ 1.522 _{-0.349} ^{+0.971}$ \\
 & $-21.12$ &  8 & $ 0.945 _{-0.276} ^{+0.768}$ \\
 & $-21.62$ &  3 & $ 0.357 _{-0.170} ^{+0.474}$ \\

\cutinhead{Photometric redshift samples}

\multirow{11}{*}{$1.4<z<1.9$}
 & $-16.19$ & 22 & $14.897 _{-2.692} ^{+7.485}$ \\
 & $-16.69$ & 26 & $10.961 _{-1.789} ^{+4.975}$ \\
 & $-17.19$ & 32 & $12.028 _{-1.764} ^{+4.904}$ \\
 & $-17.69$ & 30 & $ 9.588 _{-1.451} ^{+4.034}$ \\
 & $-18.19$ & 23 & $ 6.649 _{-1.147} ^{+3.189}$ \\
 & $-18.69$ & 25 & $ 7.112 _{-1.177} ^{+3.272}$ \\
 & $-19.19$ & 20 & $ 5.579 _{-1.032} ^{+2.868}$ \\
 & $-19.69$ & 13 & $ 3.606 _{-0.827} ^{+2.299}$ \\
 & $-20.19$ &  7 & $ 1.921 _{-0.600} ^{+1.670}$ \\
 & $-20.69$ &  3 & $ 0.814 _{-0.389} ^{+1.081}$ \\
 & $-21.19$ &  1 & $ 0.270 _{-0.223} ^{+0.621}$ \\
\hline 
 
\multirow{10}{*}{$1.8<z<2.6$}
 & $-16.55$ & 28 & $10.051 _{-1.640} ^{+4.560}$ \\
 & $-17.05$ & 44 & $11.191 _{-1.419} ^{+3.946}$ \\
 & $-17.55$ & 42 & $ 8.255 _{-1.059} ^{+2.945}$ \\
 & $-18.05$ & 40 & $ 6.873 _{-0.899} ^{+2.500}$ \\
 & $-18.55$ & 29 & $ 4.958 _{-0.761} ^{+2.117}$ \\
 & $-19.05$ & 24 & $ 4.173 _{-0.705} ^{+1.960}$ \\
 & $-19.55$ & 18 & $ 3.101 _{-0.605} ^{+1.682}$ \\
 & $-20.05$ &  7 & $ 1.217 _{-0.380} ^{+1.058}$ \\
 & $-20.55$ &  5 & $ 0.874 _{-0.323} ^{+0.899}$ \\
 & $-21.05$ &  1 & $ 0.175 _{-0.145} ^{+0.403}$ \\
\hline 
 
\multirow{10}{*}{$2.4<z<3.6$}
 & $-17.12$ & 70 & $ 9.580 _{-0.998} ^{+2.776}$ \\
 & $-17.62$ & 87 & $ 9.451 _{-0.851} ^{+2.367}$ \\
 & $-18.12$ & 59 & $ 5.413 _{-0.584} ^{+1.623}$ \\
 & $-18.62$ & 70 & $ 6.472 _{-0.640} ^{+1.780}$ \\
 & $-19.12$ & 48 & $ 4.142 _{-0.495} ^{+1.376}$ \\
 & $-19.62$ & 38 & $ 3.162 _{-0.424} ^{+1.180}$ \\
 & $-20.12$ & 17 & $ 1.399 _{-0.280} ^{+0.780}$ \\
 & $-20.62$ & 14 & $ 1.152 _{-0.254} ^{+0.708}$ \\
 & $-21.12$ &  5 & $ 0.408 _{-0.151} ^{+0.420}$ \\
 & $-21.62$ &  2 & $ 0.163 _{-0.095} ^{+0.265}$
 
\enddata
\label{tab:LFdata2}
\tablenotetext{a}{ The LF fitting is done using the modified MLE technique which does not require binning and consequently, does not use these numbers explicitly. These numbers are provided for ease of plotting of our data.}
\tablenotetext{b}{Raw number counts in the luminosity bins.}
\tablenotetext{c}{Completeness corrected number densities in the luminosity bins.}
\end{deluxetable}

\end{document}